\documentclass[aps,10pt,prd,notitlepage,showpacs,nofootinbib,superscriptaddress, compressed]{revtex4-1}

\pdfoutput=1 % if your are submitting a pdflatex (i.e. if you have
\usepackage[utf8]{inputenc}
\usepackage[english]{babel}
\usepackage{amsmath}
\usepackage{graphicx}
\usepackage{dcolumn}
\usepackage{pbox}
\usepackage{amssymb}
\usepackage{epsfig}
\usepackage[dvipsnames]{xcolor}
\usepackage{slashed}
\usepackage{amssymb}
\usepackage{mathrsfs}
\usepackage{rotating}
\usepackage{color}
\usepackage[normalem]{ulem}
\usepackage{multirow}
\usepackage[font=small]{caption}
\usepackage[font=small]{subcaption}
\usepackage{url}
\usepackage{braket}
\definecolor{DarkBlue}{rgb}{0.7, 0.4, 1} 
\definecolor{Blue}{rgb}{0, 0.8, 0} 
\definecolor{MyLightBlue}{rgb}{0.5,0.7,1.9}
\definecolor{MyGreen}{rgb}{0.0,0.2, 0.0}
\definecolor{MyBrickRed}{rgb}{0, 0.5, 0.2}
\RequirePackage{hyperref}
\hypersetup{colorlinks, citecolor=Blue,linkcolor=DarkBlue, urlcolor=Green}

\newcommand{\bea}{\begin{eqnarray}}
\newcommand{\eea}{\end{eqnarray}}

\makeatletter

\makeatletter
\renewcommand\@makecaption[2]{%
  \par
  
  \vskip\abovecaptionskip
  \begingroup
  
   \small\rmfamily
    \begingroup
     \samepage
     \flushing
     \let\footnote\@footnotemark@gobble
     \@make@capt@title{#1}{#2}\par
    \endgroup
  \endgroup
  \vskip\belowcaptionskip
}
\makeatother

\DeclareUnicodeCharacter{2212}{-}
\begin{document}
%%%%%%%%%%%%%%%%%%%%%%%%%%%%%%
\title{$Z^\prime$ induced forward dominant processes in $\mu$TRISTAN experiment}
%%%%%%%%%%%%%%%%%%%%%%%%%%%%%%    
\author{Arindam Das}
\email{adas@particle.sci.hokudai.ac.jp}
\affiliation{Institute for the Advancement of Higher Education, Hokkaido University, Sapporo 060-0817, Japan}
\affiliation{Department of Physics, Hokkaido University, Sapporo 060-0810, Japan}
\author{Yuta Orikasa}\email{yuta.orikasa@utef.cvut.cz}
\affiliation{Institute of Experimental and Applied Physics, Czech Technical University in Prague, Husova 240/5, 110 00 Prague 1, Czech Republic}
%%%%%%%%%%%%%%%%%%%%%%%%%%%%%%
\begin{abstract}
General $U(1)$ extension of the Standard Model (SM) is a well motivated beyond the Standard Model(BSM) scenario where three generations of right handed neutrinos (RHNs) are introduced to cancel gauge and mixed gauge-gravity anomalies. After the $U(1)_X$ is broken, RHNs participate in the seesaw mechanism to generate light neutrino masses satisfying neutrino oscillation data. In addition to that, a neutral gauge boson $Z^\prime$ is evolved which interacts with the left and right handed fermions differently manifesting chiral nature of the model which could be probed in future collider experiments. As a result, if we consider $\mu^+ e^-$ and $\mu^+ \mu^+$ collisions in $\mu$TRISTAN experiment $Z^\prime$ mediated $2\to2$ scattering will appear in $t-$ and $u-$channels depending on the initial and final states being accompanied by the photon and $Z$ mediated interactions. This will result well motivated resulting forward dominant scenarios giving rise to sizable left-right asymmetry. Estimating constraints on general $U(1)$ coupling from LEP-II and LHC for different $U(1)_X$ charges, we calculate differential and integrated scattering cross section and left-right asymmetry for $\mu^+ e^- \to \mu^+ e^-$ and $\mu^+ \mu^+ \to \mu^+ \mu^+$ processes which could be probed at $\mu$TRISTAN experiment further enlightening the interaction between $Z^\prime$ and charged leptons and the $U(1)_X$ breaking scale. 
\noindent 
\end{abstract}

\maketitle
%%%%%%%%%%%%%%%%%%%%%%%%%%%%%%%%%%%%%%%%%%%%%%%%%%
\section{Introduction}
%%%%%%%%main text%%%%%%%%%%%%%%%%%%%%%%%%%%%%%%%%%
Standard Model (SM) can not explain the observation of tiny neutrino mass and flavor mixing, dark matter relic density and matter-antimatter asymmetry in the universe \cite{Zyla:2020zbs} which allow us to propose theories beyond the SM (BSM). An interesting but simple extension of the SM which caters these aspects could be realized in the form $U(1)$ extension where three generations of right-handed neutrinos (RHNs) could be involved to explain the origin of tiny neutrino mass through seesaw mechanism along with an SM-singlet scalar field, responsible for the neutrino mass generation mechanism. Such scenarios are associated with an additional neutral gauge boson, commonly known as $Z^\prime$ which has very rich phenomenological aspects \cite{Leike:1998wr,Langacker:2008yv}. There are many ultraviolet-complete scenarios where $Z^\prime$ boson arises naturally, for example, Left-Right symmetric models 
\cite{Pati:1974yy,Mohapatra:1974gc,Senjanovic:1975rk}, grand unification \cite{Georgi:1974my, Fritzsch:1974nn} based on SO$(10)$
and E$_6$~\cite{Gursey:1975ki,Achiman:1978vg} theories, respectively. There is another interesting example of where $U(1)$ extension of the SM comes up with neutrino mass and aspects of dark matter and matter-antimatter asymmetry is commonly known as B$-$L  (baryon minus lepton) scenario \cite{Davidson:1978pm,Davidson:1979wr,Marshak:1979fm,
Mohapatra:1980qe} where $Z^\prime$ arises naturally and acquires mass after the B$-$L symmetry breaking. $Z^\prime$ induced scenarios are under scanner for a long period of time starting from LEP~\cite{Schael:2013ita} and  Tevatron~\cite{D0:2010kuq, CDF:2011nxq}, however, stringent bounds on the $Z^\prime$ mass and $U(1)$ gauge coupling comes from dilepton searches \cite{Aad:2019fac,CMS:2019tbu} at the LHC superseding low-energy electroweak constraints~\cite{Li:2009xh}. 
 
In this paper we investigate a general $U(1)_X$ extension of the SM which includes three generations of SM-singlet RHNs and an SM-singlet scalar which helps to break the general $U(1)_X$ symmetry by which seesaw scenario is induced to generate tiny neutrino mass followed by the generation of mass of the BSM neutral gauge boson $Z^\prime$. In our analysis we consider the situation where $U(1)_X$ charge assignment for the fermions do not depend on the generations. Reproducing the SM Yukawa interactions in the model structure we find that the $U(1)_X$ charge assignment of the SM charged fermions can be identified as the linear combination of charges of $U(1)_Y$ in SM and $U(1)_{B-L}$ gauge groups \cite{Appelquist:2002mw,Coriano:2014mpa,Das:2016zue,Das:2021esm,KA:2023dyz}. Hence the $U(1)_X$ scenario is a generalization of the $U(1)_{B-L}$ extension of the SM. Due to the gauge structure of the model we find that after solving the gauge and mixed gauge-gravity anomaly equations we find that $Z^\prime$ interacts differently with left and right handed fermions manifesting chiral nature of the model. In this paper we will focus on this property of the general $U(1)_X$ extension of the SM in the context of a muon-positron collider and multi-TeV same sign muon collider \cite{Heusch:1995yw,Arbuzov:2021oxs,Bondarenko:2021eni,Hamada:2022uyn,Hamada:2022mua,Fridell:2023gjx,Lichtenstein:2023iut,Dev:2023nha,Celada:2023oji,Goudelis:2023yni}
at $\mu$TRISTAN experiment where ultra-cold anti-muon technology can be used from J-PARC \cite{Abe:2019thb} experiment. In this experiment $\mu^+ e^-$ collision could take place at the center of mass neregy $\sqrt{s}=346$ GeV where the energy of the $\mu^+$ beam is $E_\mu=1$ TeV and that of the $e^-$ beam is $E_e=30$ GeV. Therefore center of mass energy of $\mu^+ e^-$ collider can be derived as $\sqrt{s}=2\sqrt{E_\mu E_e}$ and that could attain a luminosity of 1 ab$^{-1}$. This could be upgraded to a $\mu^+ \mu^+$ collider where center of mass energy could reach at $\sqrt{s}=2$ TeV or higher where muon beam energy could be at least $1$ TeV. This collider could attain a luminosity of $100$ fb$^{-1}$. Such machines could be potentially rich for a variety of phenomenological aspects involving physics beyond the SM apart from being a Higgs factory \cite{Bossi:2020yne,Lu:2020dkx,Cheung:2021iev,Yang:2020rjt,Liu:2021jyc,Yang:2023ojm,Das:2022mmh,Lichtenstein:2023vza}. 

Due to the presence of the $Z^\prime$ boson under the general $U(1)_X$ extension, we study $\mu^+ e^- \to \mu^+ e^-$ process mediated by $Z,~\gamma$ and $Z^\prime$ in the $t-$channel. This is process is forward dominant and therefore at $\mu$TRISTAN collider left-right asymmetry $(\mathcal{A}_{\rm{LR}})$ will be an interesting variable to study the influence of $Z^\prime$. On the other hand in the case of $\mu^+ \mu^+$ collision we study $\mathcal{A}_{\rm{LR}}$ from the $Z,~\gamma$ and $Z^\prime$ mediated $\mu^+ \mu^+ \to \mu^+ \mu^+ $ in $t-$ and $u-$channels. We involve corresponding interference terms in our analysis depending on the nature of the collision. In both the cases we first estimate total scattering cross section and estimate its deviation from the SM processes mediated by $Z$ and $\gamma$. In the case of $U(1)_X$ scenario, there will be additional $Z^\prime$ mediated processes and the corresponding interferences will affect these processes. Therefore we estimate the corresponding total scattering cross sections and their deviations from the SM. In the case of general $U(1)_X$ scenario we can vary the $U(1)_X$ charges to estimate the deviations in scattering cross section and $\mathcal{A}_{\rm{LR}}$. The effect of $U(1)_X$ charges will appear through the $Z^\prime$ vertices with electron and muon. As a result chiral nature of the model could be investigated in $\mu$TRISTAN experiment. 

The paper is organized as follows. We discuss the general $U(1)_X$ model in Sec.~\ref{secII}. We calculate total scattering cross sections for $\mu^+ e^- \to \mu^+ e^-$ and $\mu^+ \mu^+ \to \mu^+ \mu^+$ processes and we also estimate differential and integrated LR asymmetry in Sec.~\ref{secIII}. Finally we conclude the paper in Sec.~\ref{secV}.

%%%%%%%%%%%%%%%%%%%%%%%%%%%%%%%%%%%%%%%%%%%%%%%%%
\section{$U(1)$ extended scenarios and $Z^\prime$ interactions}
\label{secII}
%%%%%%%%%%%%%%%%%%%%%%%%%%%%%%%%%%%%%%%%%%%%%%%%%
\begin{table}[t]
\begin{center}
\small
\begin{tabular}{|c|c|c|c|c|c|c|c|c|} \hline
Gauge group & $q_{L}^i$ & $u_{R}^i$ & $d_{R}^i$ & $\ell_{L}^i$ & $e_{R}^i$ & $N_{R}^i$ & $H$ & $ \Phi $ \\ \hline
S$U(3)_{{C}}$ & ${\bf 3}$ & ${\bf 3}$ & ${\bf 3}$ & ${\bf 1}$ & ${\bf 1}$ & ${\bf 1}$ & ${\bf 1}$ & ${\bf 1}$ \\ \hline 
S$U(2)_{{L}}$ & ${\bf 2}$ & ${\bf 1}$ & ${\bf 1}$ & ${\bf 2}$ & ${\bf 1}$ & ${\bf 1}$ & ${\bf 2}$ & ${\bf 1}$ \\ \hline 
$U(1)_{{Y}}$ & $1/6$ & $2/3$ & $-1/3$ & $-1/2$ & $-1$ & $0$ & $1/2$ & $0$ \\ \hline
$U(1)_X$ 
%&$x_q^\prime=$&$x_u^\prime=$&$x_d^\prime=$&$x_\ell^\prime=$&$x_e^\prime=$&$x_\nu^\prime=$&$- \frac{x_H}{2}=$&$2x_\Phi=$\\ 
 & $\tilde{x}_q= \frac{1}{6}x_H + \frac{1}{3}x_\Phi$ & $\tilde{x}_u=\frac{2}{3}x_H+\frac{1}{3}x_\Phi$ & $\tilde{x}_d=-\frac{1}{3}x_H+\frac{1}{3}x_\Phi$ & $\tilde{x}_{\ell}=-\frac{1}{2}x_H-x_\Phi$ & $\tilde{x}_e=-x_H-x_\Phi$ & $\tilde{x}_e=-x_\Phi$ & $-\frac{x_H}{2}$ & $2x_\Phi$ \\ \hline
\end{tabular}
\caption{Particle content of  the minimal $U(1)_X$ model where $i(=1, 2, 3)$ is generation index.}
\label{tab1}
\end{center}
\end{table}
We consider a general $U(1)_X$ extension of the SM following the gauge structure $SU(3)_C\otimes SU(2)_L\otimes U(1)_Y\otimes$ $U(1)_X$ and the particle content is given in Tab.~\ref{tab1}. The model consists of three generations of the RHNs ($N_R^i$) and an SM-singlet BSM scalar $(\Phi)$ apart from the SM particles. The RHNs are induced to cancel gauge and mixed gauge-gravity anomalies following the corresponding anomaly conditions given below:
\begin{align}
{\rm U}(1)_X \otimes \left[ {\rm SU}(3)_C \right]^2&\ :&
			2\tilde{x}_q - \tilde{x}_u - \tilde{x}_d &\ =\  0~, \nonumber \\
{\rm U}(1)_X \otimes \left[ {\rm SU}(2)_L \right]^2&\ :&
			3\tilde{x}_q + \tilde{x}_\ell &\ =\  0~, \nonumber \\
{\rm U}(1)_X \otimes \left[ {\rm U}(1)_Y \right]^2&\ :&
			\tilde{x}_q - 8\tilde{x}_u - 2\tilde{x}_d + 3\tilde{x}_\ell - 6\tilde{x}_e &\ =\  0~, \nonumber \\
\left[ {\rm U}(1)_X \right]^2 \otimes {\rm U}(1)_Y &\ :&
			{\tilde{x}_q}^2 - {\tilde{2x}_u}^2 + {\tilde{x}_d}^2 - {\tilde{x}_\ell}^2 + {\tilde{x}_e}^2 &\ =\  0~, \nonumber \\
\left[ {\rm U}(1)_X \right]^3&\ :&
			{6\tilde{x}_q}^3 - {3\tilde{x}_u}^3 - {3 \tilde{x}_d}^3 + {2\tilde{x}_\ell}^3 - {\tilde{x}_\nu}^3 - {\tilde{x}_e}^3 &\ =\  0~, \nonumber \\
{\rm U}(1)_X \otimes \left[ {\rm grav.} \right]^2&\ :&
			6\tilde{x}_q - 3\tilde{x}_u - 3 \tilde{x}_d + 2\tilde{x}_\ell - \tilde{x}_\nu - \tilde{x}_e &\ =\  0~, 
\label{anom-f-1}
\end{align}
respectively. We write down the Yukawa interactions in the model involving SM and BSM particles following the gauge structure $\mathcal{G}_{\rm SM} \otimes$ U$(1)_X$ as
\begin{equation}
{\cal L}^{\rm Yukawa} = - Y_u^{\alpha \beta} \overline{q_L^\alpha} H u_R^\beta
                                - Y_d^{\alpha \beta} \overline{q_L^\alpha} \tilde{H} d_R^\beta
				 - Y_e^{\alpha \beta} \overline{\ell_L^\alpha} \tilde{H} e_R^\beta
				- Y_\nu^{\alpha \beta} \overline{\ell_L^\alpha} H N_R^\beta- Y_N^\alpha \Phi \overline{(N_R^\alpha)^c} N_R^\alpha + {\rm H.c.}~,
\label{LYk1}   
\end{equation}
where SM Higgs doublet filed can be denoted by $H$ following the transformation rule $\tilde{H}$ following $i  \tau^2 H^*$ where $\tau^2$ is the second Pauli matrix. Now following charge neutrality in the context of Yukawa interactions given in Eq.~(\ref{LYk1}) we write the following relations between the $U(1)_X$ charges of the particles as 
\begin{eqnarray}
-\frac{1}{2} x_H^{} &=& - \tilde{x}_q + \tilde{x}_u \ =\  \tilde{x}_q - \tilde{x}_d \ =\  \tilde{x}_\ell - \tilde{x}_e=\  - \tilde{x}_\ell + \tilde{x}_\nu, \nonumber \\
2 x_\Phi^{}	&=& - 2 \tilde{x}_\nu~. 
\label{Yuk}
\end{eqnarray} 
Solving Eqs.~\eqref{anom-f-1} and \eqref{Yuk} we obtain the $U(1)_X$ charges of the particles in the model in terms of the charges of the scalar fields $x_H$ and $x_\Phi$.
The charges are given in the Tab.~\ref{tab1}. We find that after anomaly cancellation, the charges are written as a linear combination of $U(1)_Y$ and $U(1)_{\rm B-L}$ charges.
Hence note that left and right handed fermions are differently charges under the $U(1)_X$ gauge group leading to chiral nature of the model while these particles interact with the neutral BSM gauge boson $Z^\prime$. In our analysis we fix $x_\Phi=1$ without the loss of generality and vary $x_H$ showing interesting properties. If we fix $x_H=-2$, we find that left handed fermions do not interact with $Z^\prime$, however, right handed fermions do. This is $U(1)_R$ scenario while for $x_H=0$ we find that left and right handed fermions interact equally with $Z^\prime$ reproducing a vector-like scenario commonly known as B$-$L model. In addition to that we fix $x_H=-1$ when $e_R$ does not interact with $Z^\prime$, however, $\ell_L$ does. Similarly for $x_H=1$, the $U()1)_X$ charge of $d_R$ becomes zero disallowing its interaction with $Z^\prime$. For simplicity we fix $x_H=2$ where all left and right handed fermions interact with $Z^\prime$, however, their $U(1)_X$ charges are different manifesting the chiral nature of the model. In our analysis we will study the left right asymmetry from the $\mu^+ e^- \to \mu^+ e^-$ and $\mu^+ \mu^+ \to \mu^+ \mu^+$ forward dominant processes, therefore we confine ourselves in the chiral scenarios only. The renormalizable scalar potential of this model can be written as
\begin{align}
  V \ = \ m_H^2(H^\dag H) + \lambda_H^{} (H^\dag H)^2 + m_\Phi^2 (\Phi^\dag \Phi) + \lambda_\Phi^{} (\Phi^\dag \Phi)^2 + \lambda_{\rm mix} (H^\dag H)(\Phi^\dag \Phi)~,
\end{align}
where $H$ and $\Phi$ can be separately approximated in the analysis of scalar potential considering $\lambda_{\rm mix}$ to be very small. After the $U(1)_X$ symmetry breaking and the electroweak symmetry breaking, the scalar fields develop vacuum expectation values (VEVs) as
\begin{align}
  \braket{H}=  \frac{1}{\sqrt{2}}\begin{pmatrix} v+h\\0 
  \end{pmatrix}~ \quad {\rm and}\quad 
\braket{\Phi}  =  \frac{v_\Phi^{}+\phi}{\sqrt{2}}~,
\end{align}
where $v=246$ GeV is the electroweak scale VEV at the potential minimum and $v_\Phi^{}$ is a free parameter. After the breaking of $U(1)_X$ symmetry, the neutral BSM gauge boson $Z^\prime$ acquires the following mass term as  
\begin{equation}
 M_{Z^\prime}^{}=  2 g_X^{}  v_\Phi^{}~,
\end{equation}
considering $v_\Phi^{} \gg v$ and $x_\Phi=1$ and here $g_X$ is general $U(1)_X$ coupling. Here $Z^\prime$ mass is a free parameter. 

From the Yukawa interactions given in Eq.~(\ref{LYk1}), we find that the RHNs interact with the SM-singlet scalar field $\Phi$ generates Majorana mass term for heavy neutrinos after the breaking of the general $U(1)$ symmetry. After the electroweak symmetry breaking, the Dirac mass term is generated from the interaction among the SM-singlet RHN,
SM like Higgs doublet and SM lepton doublet switching on the seesaw mechanism from
\begin{equation}
    m_{N_R^\alpha}^{} \ = \ \frac{Y^\alpha_{N}v_\Phi^{}}{\sqrt{2}} , \, \, \, \, \, \,\,\,
    m_{D}^{\alpha \beta} \  =  \ \frac{Y_{\nu}^{\alpha \beta} v}{\sqrt{2}}~
\label{mDI}
\end{equation}
which are Majorana and Dirac mass terms respectively to originate the tiny neutrino masses and flavor mixing. Hence neutrino mass matrix can be written as 
\begin{equation}
   m_\nu= \begin{pmatrix} 0&m_D^{}\\ m_D^T&m_N^{} \end{pmatrix}.~
\label{num-1}
\end{equation}
Finally diagonalizing Eq.~\ref{num-1} we obtain the light neutrino mass eigenvalues as $-m_D^{} m_N^{-1} m_D^T$. As neutrino mass generation is not the main motivation of this paper therefore we skip giving much details of it. 

The interaction between the $Z^\prime$ gauge boson and SM charged fermions can be written as 
\bea
\mathcal{L}_{\rm int} \ = \ -g_X \left(\overline{q}\gamma^\mu \tilde{x}_{q} P_L q+ \overline{u}\gamma^\mu \tilde{x}_{u}  P_R u+ \overline{d}\gamma^\mu \tilde{x}_{d}  P_R d\right) Z_\mu^\prime-g_X \left(\overline{\ell}\gamma^\mu \tilde{x}_{\ell} P_L \ell+ \overline{e}\gamma^\mu \tilde{x}_{e} P_R e\right)Z_\mu^\prime,
\label{Lag1}
\eea
where $\tilde{x}_{u}$, $\tilde{x}_d$, $\tilde{x}_\ell$ and $\tilde{x}_e$ are the corresponding charges which could be obtained from Tab.~\ref{tab1}. In our paper $Z^\prime$ interaction with charged leptons are the relevant ones. Here $P_L$ and $P_R$ are the left and right projection operators $(1\mp \gamma_5)/2$ respectively. Now we show the partial decay widths of $Z^\prime$ into different modes using Eq.~\eqref{Lag1}.
The partial decay widths of $Z^\prime$ into the SM charged fermions can be written as
\bea
\Gamma(Z^\prime \to f\bar{f}) \ = \ N_c \frac{M_{Z^\prime}}{24\pi}~\Big(g_L^f \Big[g_X, x_H, x_\Phi \Big]^2 + g_R^f \Big[g_X, x_H, x_\Phi \Big]^2\Big) \, ,
\label{2f}
\eea
where $N_c=3\ (1)$ is a color factor for the quarks (leptons) and $g_{L(R)}^f\Big[g_X, x_H, x_\Phi \Big]$ is the coupling of the $Z^\prime$ with the left (right) handed charged fermions from SM and these couplings depend on the $U(1)_X$ charges. The partial decay width of the $Z^\prime$ into light neutrinos can be written as 
\bea
\Gamma(Z^\prime \to \nu\bar{\nu}) \ = \ 3 \frac{M_{Z^\prime}}{24\pi}~g_L^\nu \Big[g_X, x_H, x_\Phi \Big]^2.
\label{2v}
\eea 
The partial decay width of the $Z^\prime$ into a pair of RHNs for three generations can be written as 
\bea
\Gamma(Z^\prime \to N N) \ = \ 3 \frac{M_{Z^\prime}}{24\pi}~g_R^N \Big[g_X, x_\Phi \Big]^2 \Big(1-4\frac{m_N^2}{M_{Z^\prime}^2}\Big)^{\frac{3}{2}} \, .
\label{2N}
\eea
However, in this analysis we assume for simplicity that the decay of the $Z^\prime$ into a pair of RHNs is kinematically disallowed because $m_N > M_{Z^\prime}$.

\begin{figure}[h]
\includegraphics[width=0.85\textwidth,angle=0]{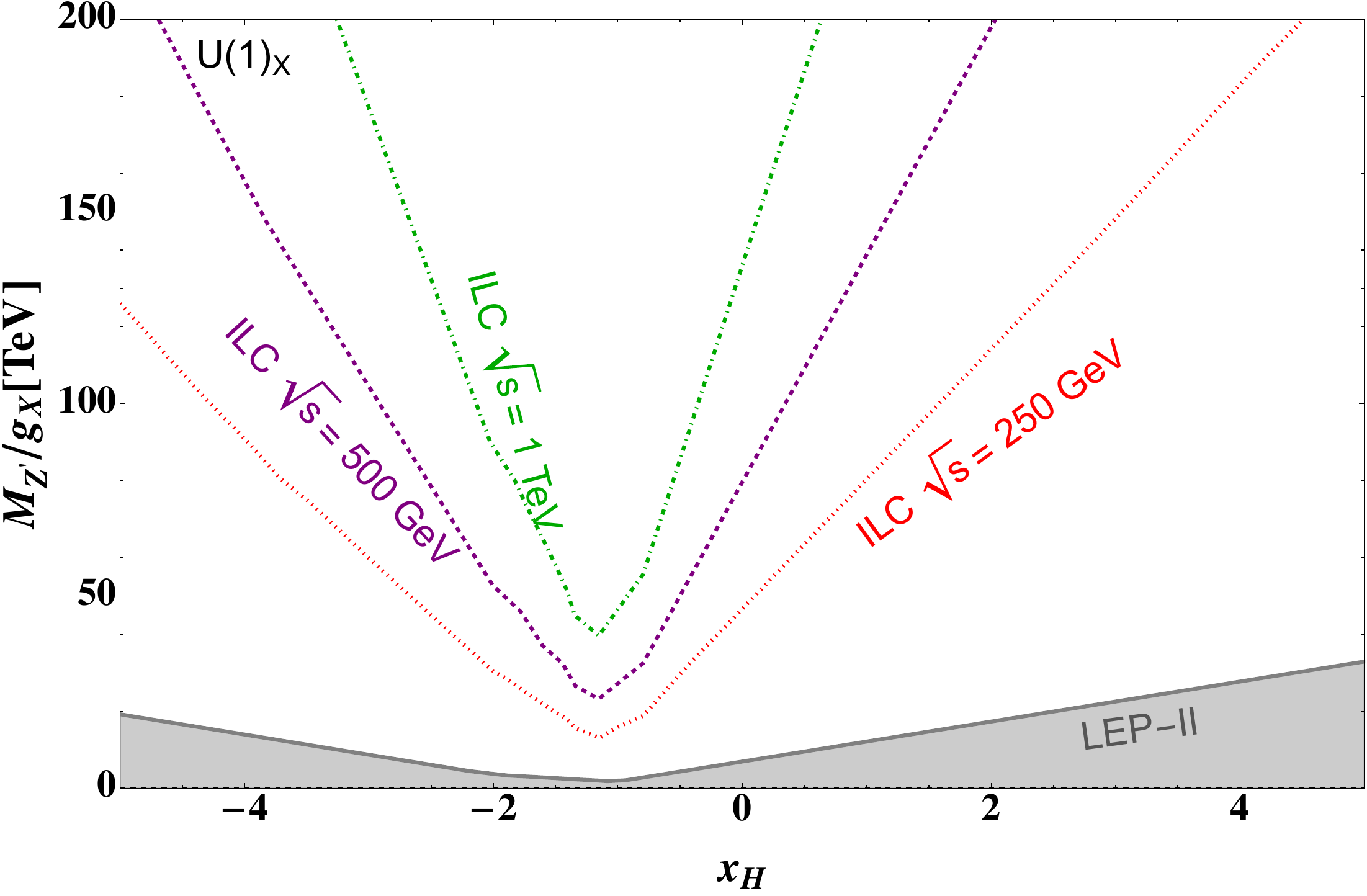}
\caption{Limits on $M_{Z^\prime}/g_X$ vs $x_H$ in $U(1)_X$ scenario for LEP-II and prospective ILC in $e^- e^+$ collision. The LEP-II limits rule out the parameter space shown by gray shaded region. Prospective limits from ILC using $e^-e^+$ collisions are shown by different colored lines without shading.} 
\label{fig1}
\end{figure}

\begin{figure}[h]
\includegraphics[width=0.497\textwidth,angle=0]{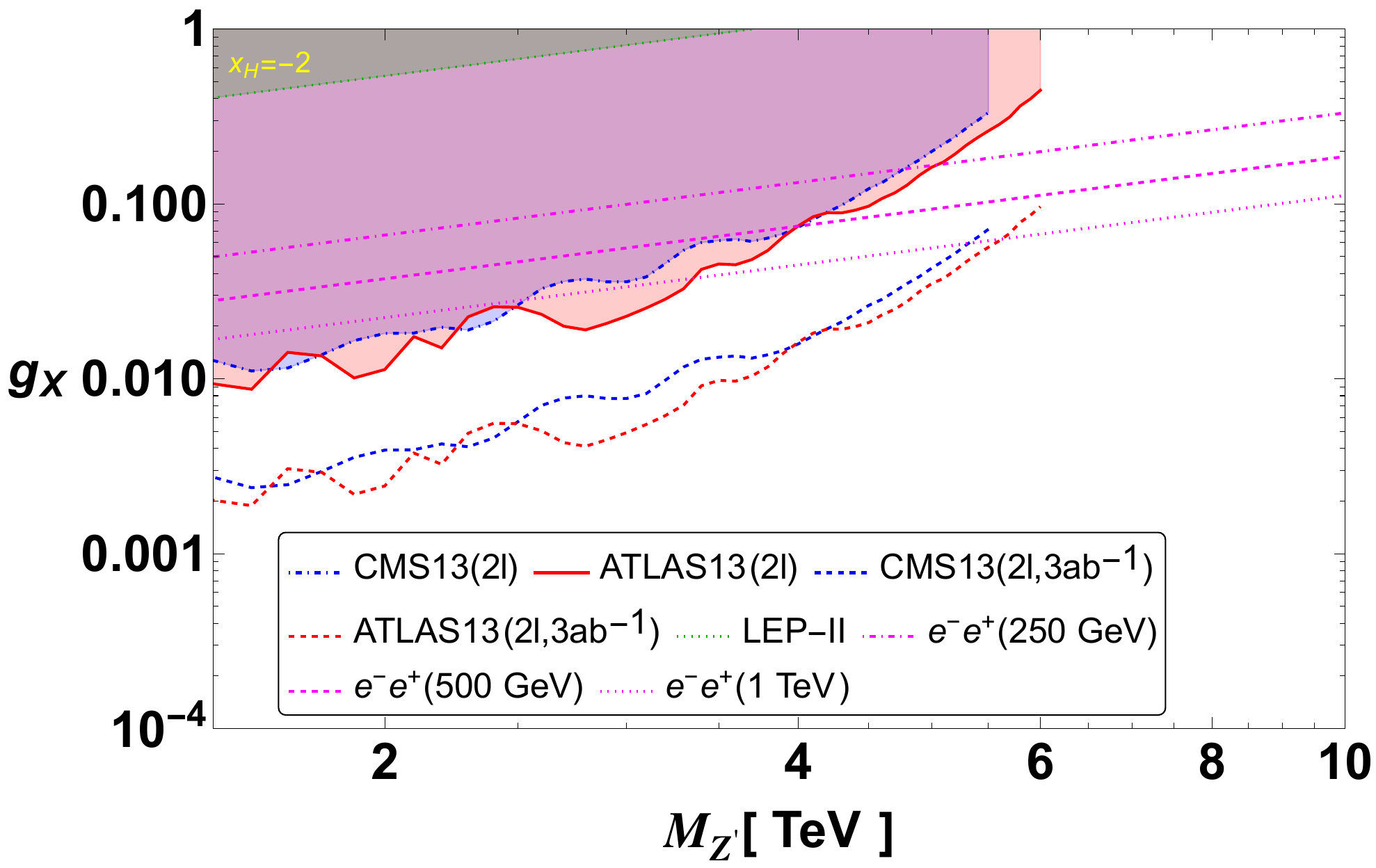}
\includegraphics[width=0.497\textwidth,angle=0]{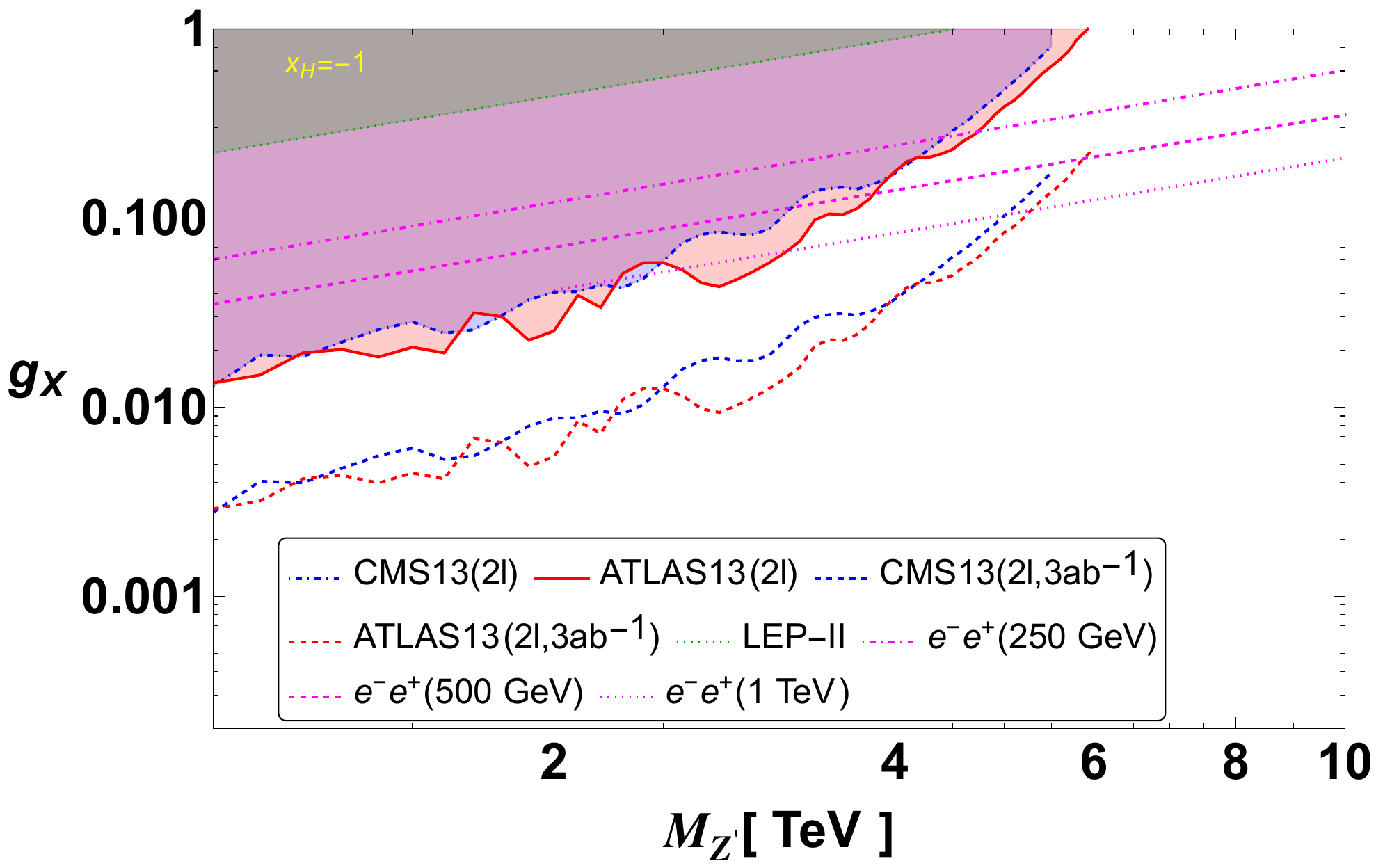}\\
\includegraphics[width=0.497\textwidth,angle=0]{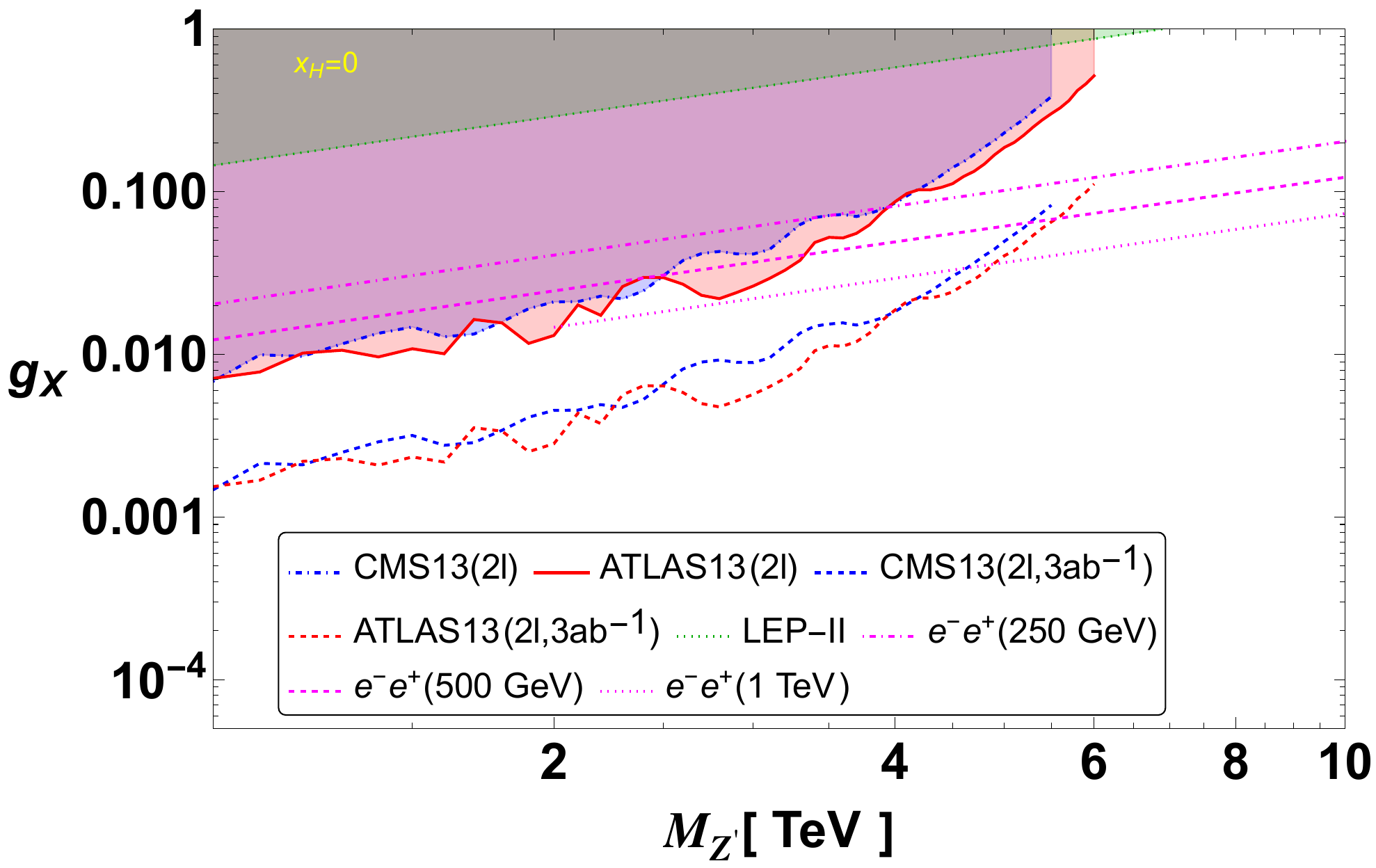}\\
\includegraphics[width=0.497\textwidth,angle=0]{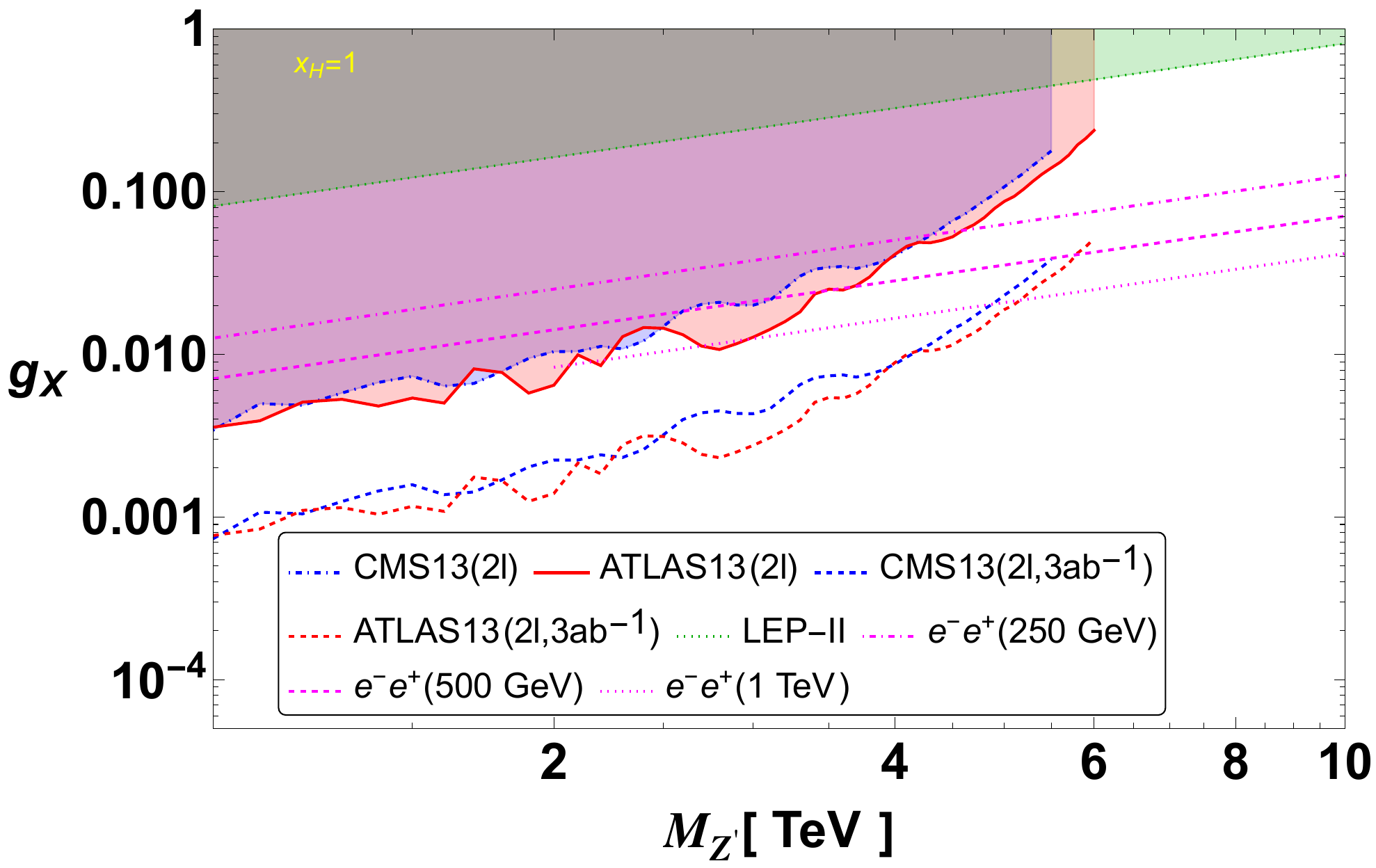}
\includegraphics[width=0.497\textwidth,angle=0]{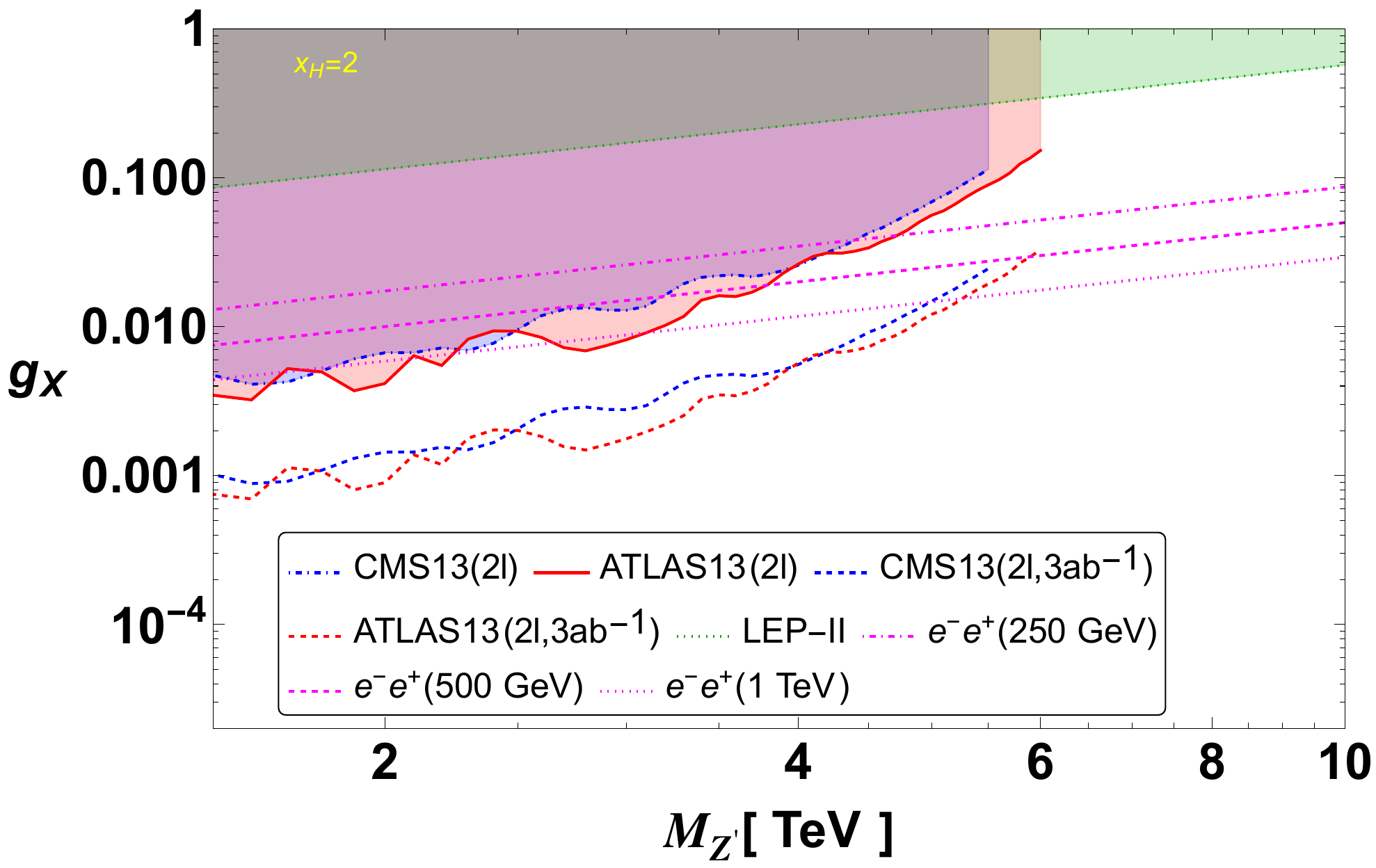}
\caption{$95\%$ upper limits on $g_X-M_{Z^\prime}$ plane for different $x_H$. The shaded regions are are ruled out by LEP-II (green), CMS (blue) and ATLAS (red) experiments respectively.The prospective limits are coming from $e^-e^+$ collisions in ILC at $\sqrt{s}=250$ GeV, $500$ GeV and $1$ TeV by magenta dot-dashed, dashed and dotted lines respectively. Prospective limits from  LHC(ATLAS and CMS) at $3$ ab$^{-1}$ luminosity shown by dashed blue and red lines.} 
\label{fig2}
\end{figure}

Now we calculate bounds on $g_X$ with respect to $M_{Z^\prime}$. We calculate the limits on model parameters from the LEP experiment first for different $x_H$ taking $M_{Z^\prime} \gg \sqrt{s}$ in consideration. Following \cite{Eichten:1983hw, LEP:2003aa,Schael:2013ita} and utilizing contact interaction for the process $e^-e^+ \to f \bar{f}$ we finally parametrize 
\bea
{\cal L}_{\rm eff} \ = \ \frac{g_X^2}{(1+\delta_{ef}) (\Lambda_{AB}^{f\pm})^2} \sum_{A,B=L,R}\eta_{AB}(\overline{e} \gamma^\mu P_A e)(\overline{f} \gamma_\mu P_B f) \, ,
\label{eq1}
\eea
where $g_X^2/4\pi$ is taken to be 1 by convention where $\delta_{ef}=1\ (0)$ for $f=e$ ($f\neq e$),  $\eta_{AB}=\pm 1$ or 0, and $\Lambda_{AB}^{f\pm}$ is considered as the scale of contact interaction. Here we consider constructive ($+$) or destructive ($-$) interference with SM processes $e^+e^-\to f\bar{f}$ \cite{Kroha:1991mn}. Following \cite{Carena:2004xs} we evaluate $Z^\prime$ exchange matrix element in $U(1)_X$ case as
\bea
\frac{g'^2}{{M_{Z^\prime}}^2-s} [\overline{e} \gamma^\mu (\tilde{x}_\ell P_L+ \tilde{x}_e P_R) e] [\overline{f} \gamma_\mu (\tilde{x}_{f_L} P_L+ \tilde{x}_{f_R} P_R) f] \, ,
\label{eq2}
\eea
where $\tilde{x}_{f_L}$ and $\tilde{x}_{f_R}$ are $U(1)_X$ charges of $f_L$ and $f_R$ respectively which can be found in Table~\ref{tab1}. Now we match Eqs.~\eqref{eq1} and \eqref{eq2} to obtain 
\bea
M_{Z^\prime}^2  \ \gtrsim \ \frac{{g^\prime}^2}{4\pi} |{x_{e_A}} x_{f_B}| (\Lambda_{AB}^{f\pm})^2 \, , 
\label{Lim}
\eea
considering $M_{Z^\prime}^2 \gg s$ where $\sqrt{s}=209$ GeV for LEP-II. Hence we estimate bounds on $M_{Z^\prime}/g_X$ from LEP using different values of  $\Lambda_{AB}^{f\pm}$ for different values of $x_H$. To do that we use $95\%$ bounds on $\Lambda_{AB}^{f\pm}$ from \cite{Schael:2013ita} for leptons and quarks taking $AB=LL,\ RR, \ LR, \ RL, \ VV$ and $AA$ assuming universality in the contact interactions. In the same way we estimate prospective limits on $M_{Z^\prime}/g_X$ for different $x_H$ at ILC with $\sqrt{s}=250$ GeV, $500$ GeV and $1$ TeV using the bounds on $\Lambda_{AB}^{f\pm}$ from \cite{Fujii:2019zll}. We show these bounds in Fig.~\ref{fig1} where we find that LEP-II puts a stringent bound on $M_{Z^\prime}/g_X$ and prospective limits from ILC could be stronger. The corresponding limits at $95\%$ C. L. for different $x_H$ are shown in Tab.~\ref{tab2}.
%%%%%%%%%%%%%%%%%%%%%%%%%%%%%%%%%%%%%%%%%
%%%%%%%%%%%%%%%%%%%%%%%%%%%%%%%%%%%%%%%
\begin{table}[t!]
\begin{center}
\begin{tabular}{|c|c|c|c|c|c|c|}
\hline
\multirow{2}{*}{Machine} & \multirow{2}{*}{$\sqrt s$} & \multicolumn{5}{|c|}{95\% CL lower limit on $M_{Z'}/g_X$ (in TeV)} \\ \cline{3-7}
& & $x_H=-2$ &  $x_H=-1$ &$x_H=0$ & $x_H=1$ & $x_H=2$ \\ \hline
LEP-II & 209 GeV & 5.0 & 2.2&7.0 &11.1 &18.0 \\ \hline
\multirow{3}{*}{ILC} & 250 GeV & 31.6 &16.3&48.2 &79.0 &113.7 \\ \cline{2-7}
& 500 GeV & 54.4 &26.3&81.6 & 139.1 &199.7 \\ \cline{2-7}
 & 1 TeV & 88.6 &47.7 &137.2 &238.2&339.2 \\ 
\hline
\end{tabular}
\end{center}
\caption{The $95\%$ CL lower limits on $M_{Z^\prime}/g_X$ in the $U(1)_X$ model from $e^+e^-\to f\bar{f}$ processes for different $x_H$ and taking the most stringent limit out of all the different channels considered there. }
\label{tab2}
\end{table}
%%%%%%%%%%%%%%%%%%%%%%%%%%%%%%%%%%%%%%%%%%%%%%%%%%%%%%%%%%%%
Finally solving $M_{Z^\prime}/g_X$ bounds for different $x_H$ and for a range of $M_{Z^\prime}$, we estimate limits on $g_X-M_{Z^\prime}$ plane for different $x_H$ and the limits are shown in Fig.~\ref{fig2} for LEP-II (green dotted) and prospective ILC(magenta dot-dashed, dashed and dotted), respectively. We calculate limits on $g_X-M_{Z^\prime}$ for different $x_H$ from the dilepton \cite{Aad:2019fac, CMS:2019tbu} searches at the LHC using CMS and ATLAS results. We first estimate dilepton signal for electron and muon involving $Z^\prime$ from the model $\sigma_{\rm Model}$ considering a trial value of $g_{\rm{Model}}$ for a given $x_H$ and different $M_{Z^\prime}$ at $\sqrt{s}=14$ TeV. Now we compare these cross sections with observed cross section from ATLAS and CMS $(\sigma_{\rm{Obs.}})$ and using 
\bea
g^\prime \ = \ \sqrt{g_{\rm{Model}}'^2 \left(\frac{\sigma_{\rm{Obs.}}}{\sigma_{\rm{Model}}}\right)}.
\label{gp}
\eea
Hence we calculate $95\%$ limits on $g_X-M_{Z^\prime}$ plane for different $x_H$. We also consider the future high-luminosity phase of the LHC (HL-LHC) at $\sqrt s=14$ TeV with $3~{\rm ab}^{-1}$ integrated luminosity and draw the projected dilepton bounds using 
\bea
g_X^{\rm projected}= g_X^{\rm current} \sqrt{\frac{139 (140) \rm{fb}^{-1}}{\mathcal{L}_{\rm projected}}}.
\eea
The corresponding limits for CMS (blue dot-dashed) and ATLAS (red solid) for $2\ell$ are shown in Fig.~\ref{fig2} for 139 fb$^{-1}$ and 140 fb$^{-1}$ luminosities. The scaled results for CMS (blue dashed) and ATLAS (red dashed) at 3 ab$^{-1}$ are also shown in the same figure. The shaded regions are ruled out by existing experiments.  
%%%%%%%%%%%%%%%%%%%%%%%%%%%%%%%%%%%%%%%%%
%%%%%%%%%%%%%%%%%%%%%%%%%%%%%%%%%%%%%%%
For further analysis we consider $M_{Z^\prime}=7.5$ TeV and $g_X=1$ for $x_H=-2$, $-1$ and $0$ where as for same $M_{Z^\prime}$ we consider $g_X=0.6$ and $0.5$ for $x_H=1$ and $x_H=2$ respectively. 
%%%%%%%%%%%%%%%%%%%%%%%%%%%%%%%%%%%%%%%%%%%%%%%%%%%%%%%%%%%%
\section{Scattering cross section and Left-Right asymmetry}
\label{secIII}
%%%%%%%%%%%%%%%%%%%%%%%%%%%%%%%%%%%%%%%%%%%%%%%%%%%%%%%%%%%%
We consider the $2\to2$ scattering in the context of $\mu-$TRISTAN experiment where we first consider the initial states as $\mu^+ e^-$ and then $\mu^+\mu^+$ where we have forward dominant processes through $t-$channel and $u-$channel depending on the final states. Therefore in case $\mu-$TRISTAN experiment to study the effect of $Z^\prime$ deviations in scattering cross sections from the SM and left-right asymmetry will be suitable observables to investigate in the following: 
%%%%%%%%%%%%%%%%%%%%%%%%%%%%%%%%%%%%%%%%%%
\subsection{$\mu^+ e^-$ collider}
%%%%%%%%%%%%%%%%%%%%%%%%%%%%%%%%%%%%%%%%%%%
\begin{figure}[h]
\includegraphics[width=0.85\textwidth,angle=0]{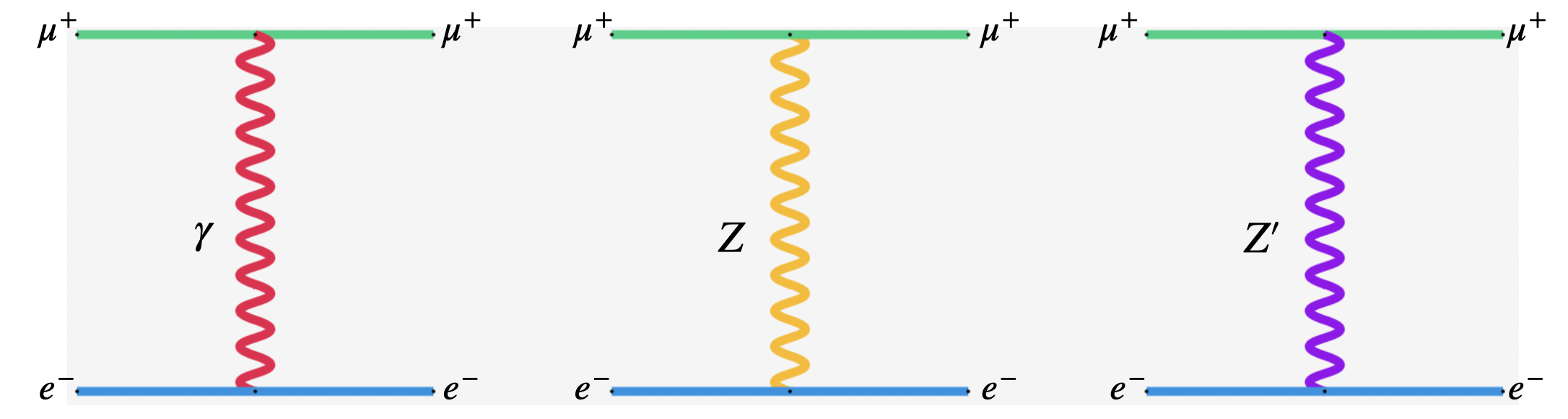}
\caption{Photon, $Z$ and $Z^\prime$ mediated $\mu^+ e^- \to \mu^+ e^-$ processes in $t-$channel at $\mu$TRISTAN experiment with $\sqrt{s}=346$ GeV where muon beam energy is 1 TeV and electron beam energy is 30 GeV.} 
\label{fig-mue}
\end{figure} 
%%%%%%%%%%%%%%
We consider the $\mu^+ e^- \to \mu^+ e^-$ scattering in the center-of-mass frame. 
Fig.~\ref{fig-mue} shows the $\mu^+ e^- \to \mu^+ e^-$ scattering processes. 
There are photon, Z and $Z^\prime$ mediated t-channel processes. We define the quantities $p_1$ as initial muon momentum, $p_2$ as initial electron momentum, $k_1$ as final muon momentum, $k_2$ as final electron momentum. The $4-$vector representation of these momenta are
$p_1 = (E_\mu, 0, 0, E_\mu)$, $p_2 = (E_e, 0, 0, -E_e)$, $k_1 = (E_1, 0, E_1 \sin\theta, E_1 \cos\theta)$ and $k_2 = (E_\mu + E_e - E_1, 0, -E_1 \sin\theta, E_\mu - E_e - E_1 \cos\theta)$, where $E_\mu$ is the initial muon energy, $E_e$ is the initial electron energy, $\theta$ is the scattering angle of the muon and $E_1=(2 E_\mu E_e)/(E_\mu + E_e - (E_\mu - E_e)\cos\theta)$. The Mandelstam variables are given by 
\begin{eqnarray}
s_e= 4 E_\mu E_e, \, \, \, t_e= - 2 E_\mu E_1 (1 - \cos\theta), \, \, \,
u_e= - 2 E_e E_1 (1 + \cos\theta). 
\end{eqnarray}
We fix the energies, $E_\mu = $1 TeV, $E_e = $ 30 GeV and the center-of-mass energy $\sqrt{s_e} = $ 346 GeV. We define the following quantities
\begin{eqnarray}
T_{RR}^e \equiv
\sum_{V=\{\gamma, Z, Z'\}} \frac{g_{L}^{V\mu} g_{R}^{Ve}}{t_e - M_V^2}, 
\,\,
T_{LL}^e \equiv 
\sum_{V=\{\gamma, Z, Z'\}} \frac{g_{R}^{V\mu} g_{L}^{Ve}}{t_e - M_V^2}, \nonumber \\
T_{RL}^e \equiv  
\sum_{V=\{\gamma, Z, Z'\}} \frac{g_{L}^{V\mu} g_{L}^{Ve}}{t_e - M_V^2},  \, \, \,
T_{LR}^e \equiv
\sum_{V=\{\gamma, Z, Z'\}} \frac{g_R^{V\mu} g_{R}^{Ve}}{t_e - M_V^2}.~~~ 
\label{Te}
\end{eqnarray}
where $M_V$ and $\Gamma_V$ are the mass and total decay width of the corresponding gauge bosons (except photon). 
%%%%%%%%%%%%%%%%%%%%%
\begin{figure}[h]
\begin{center}
\includegraphics[width=0.497\textwidth,angle=0]{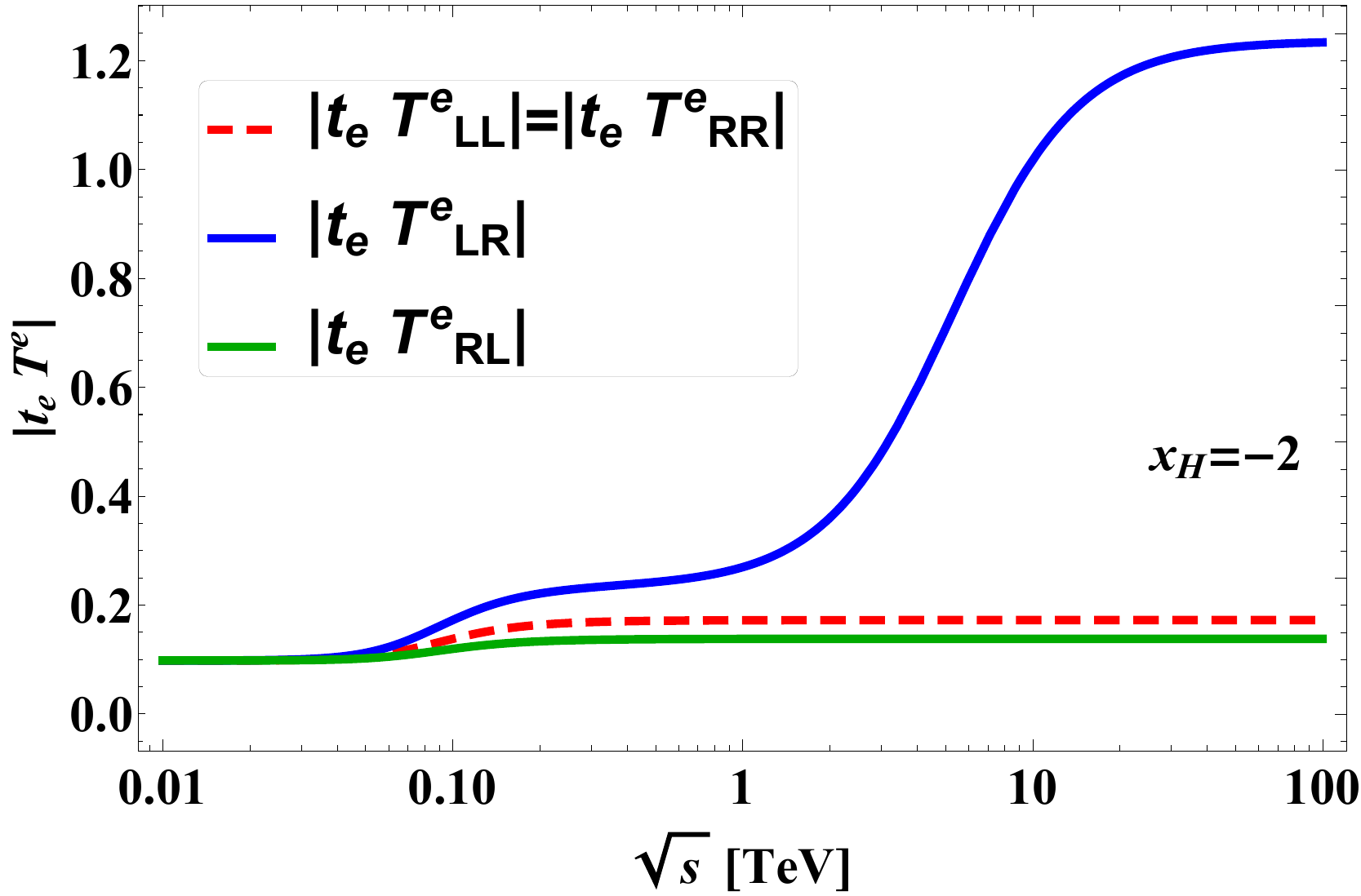}
\includegraphics[width=0.497\textwidth,angle=0]{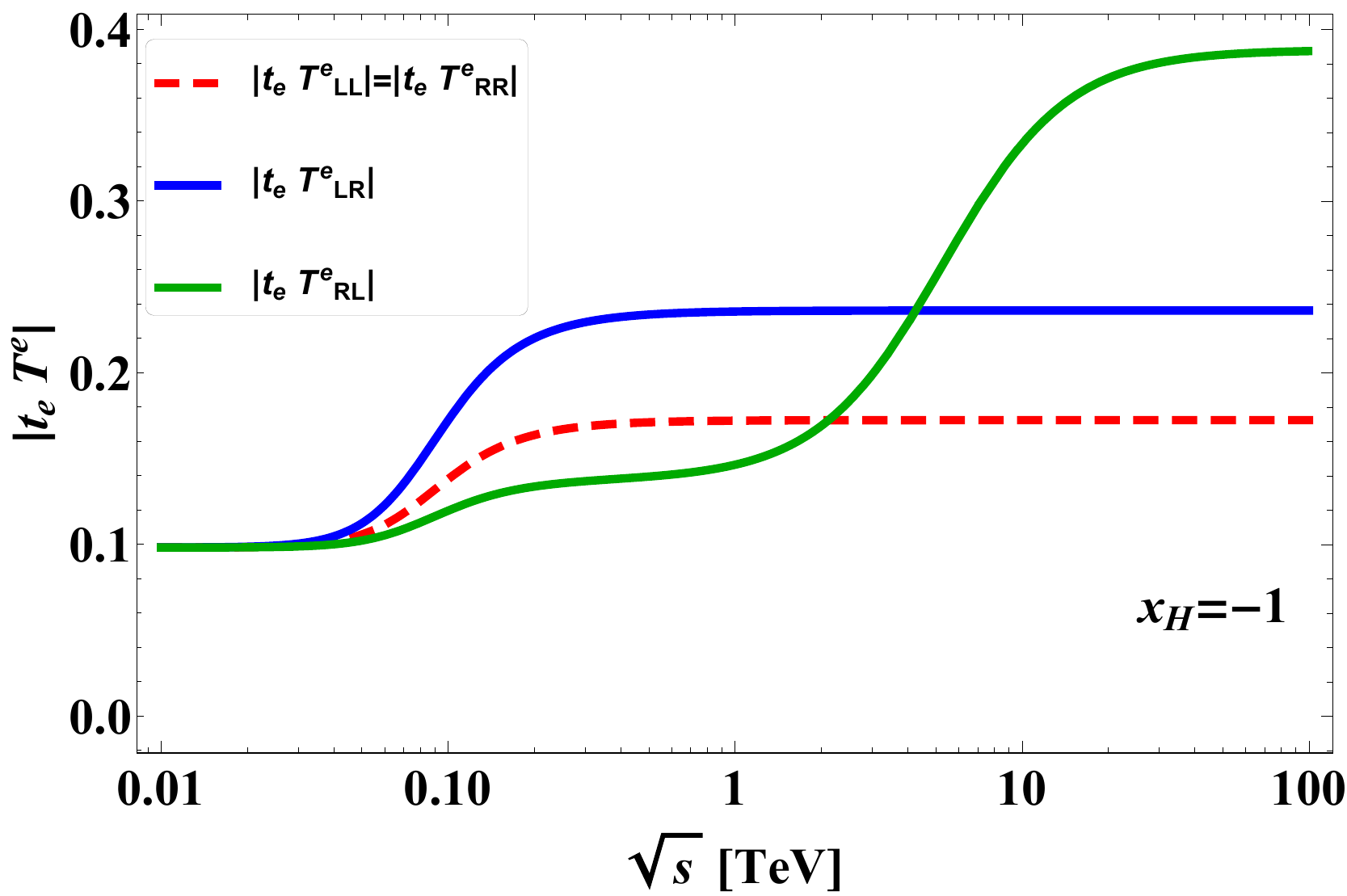}\\
\includegraphics[width=0.497\textwidth,angle=0]{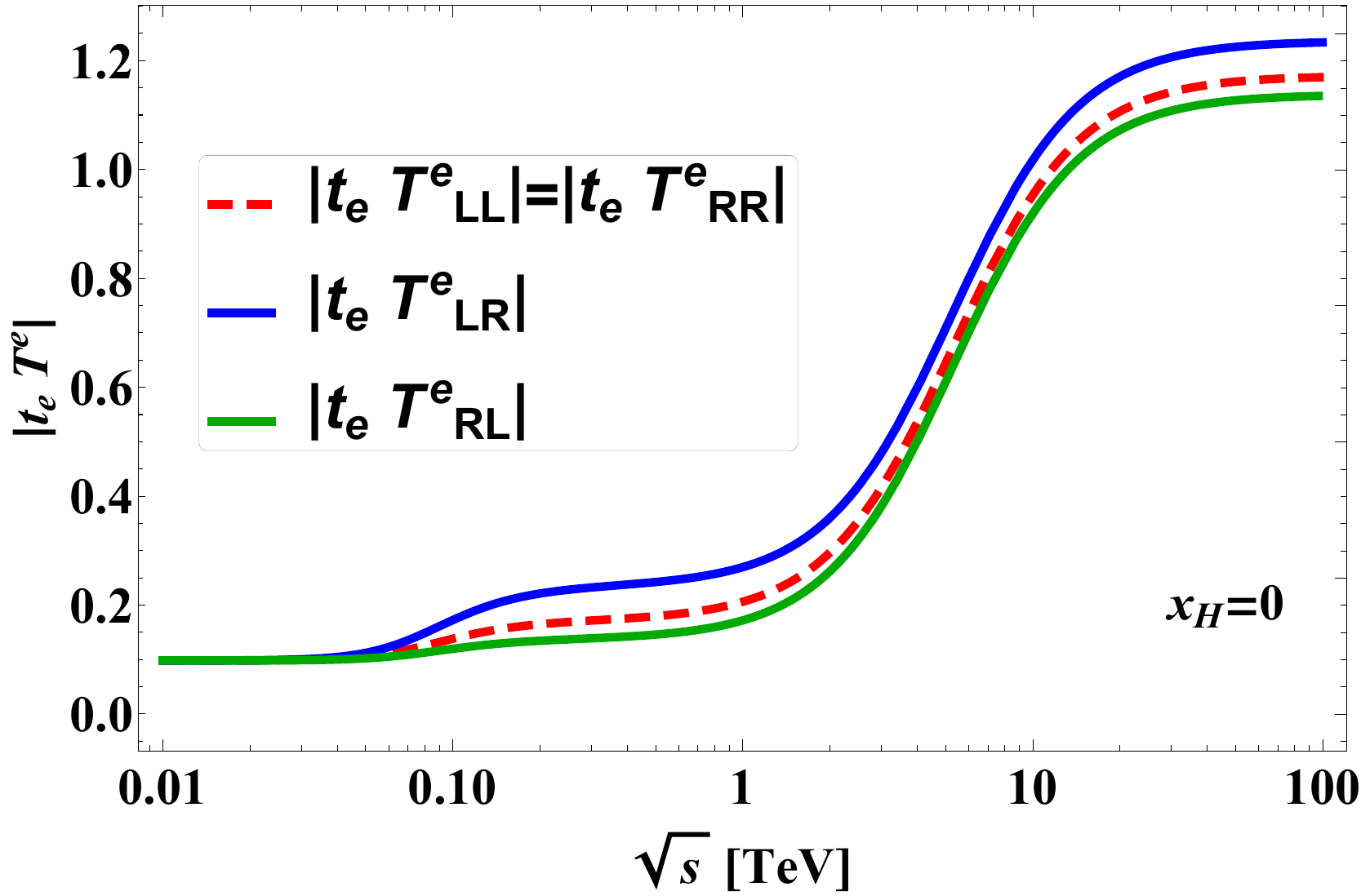}\\
\includegraphics[width=0.497\textwidth,angle=0]{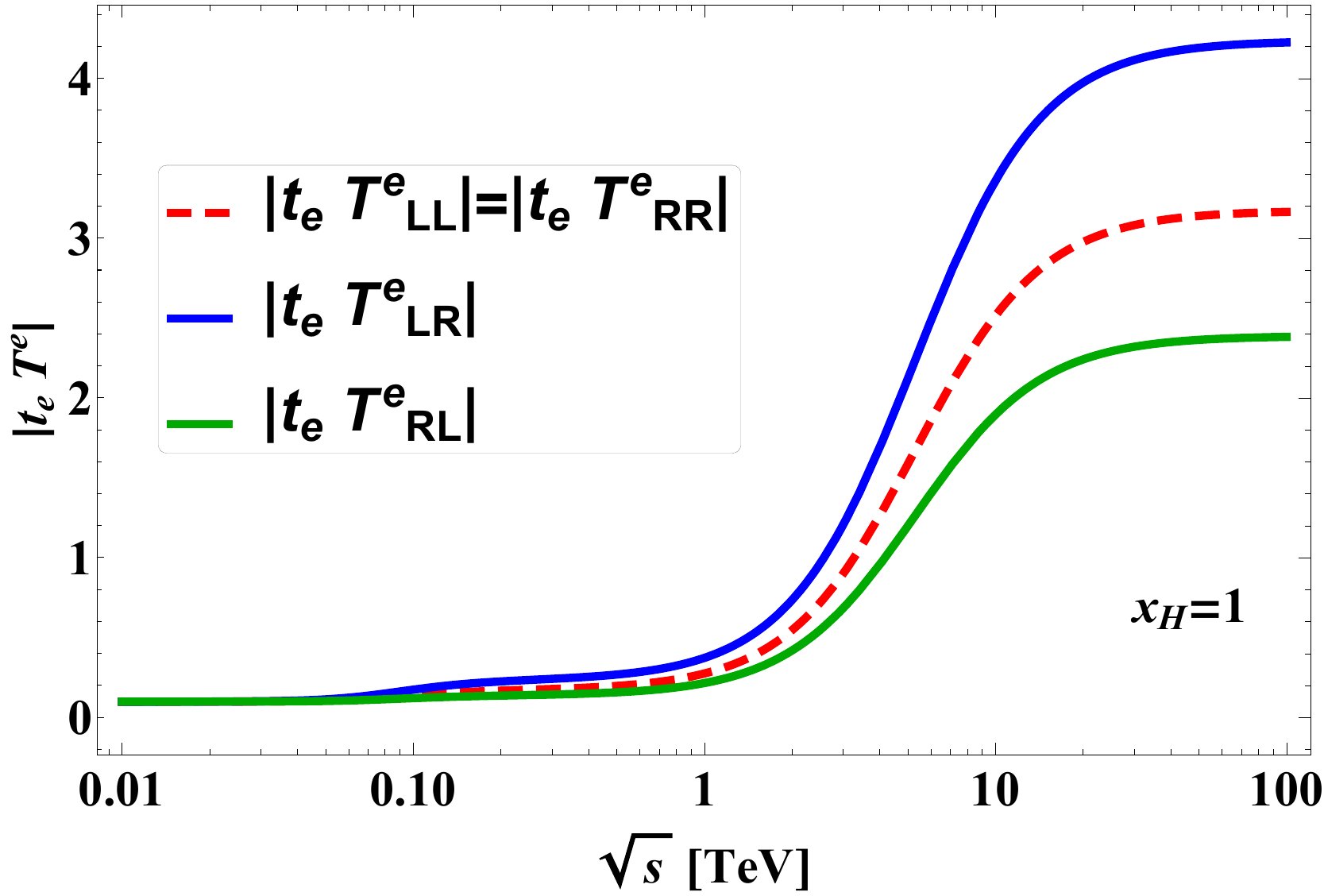}
\includegraphics[width=0.497\textwidth,angle=0]{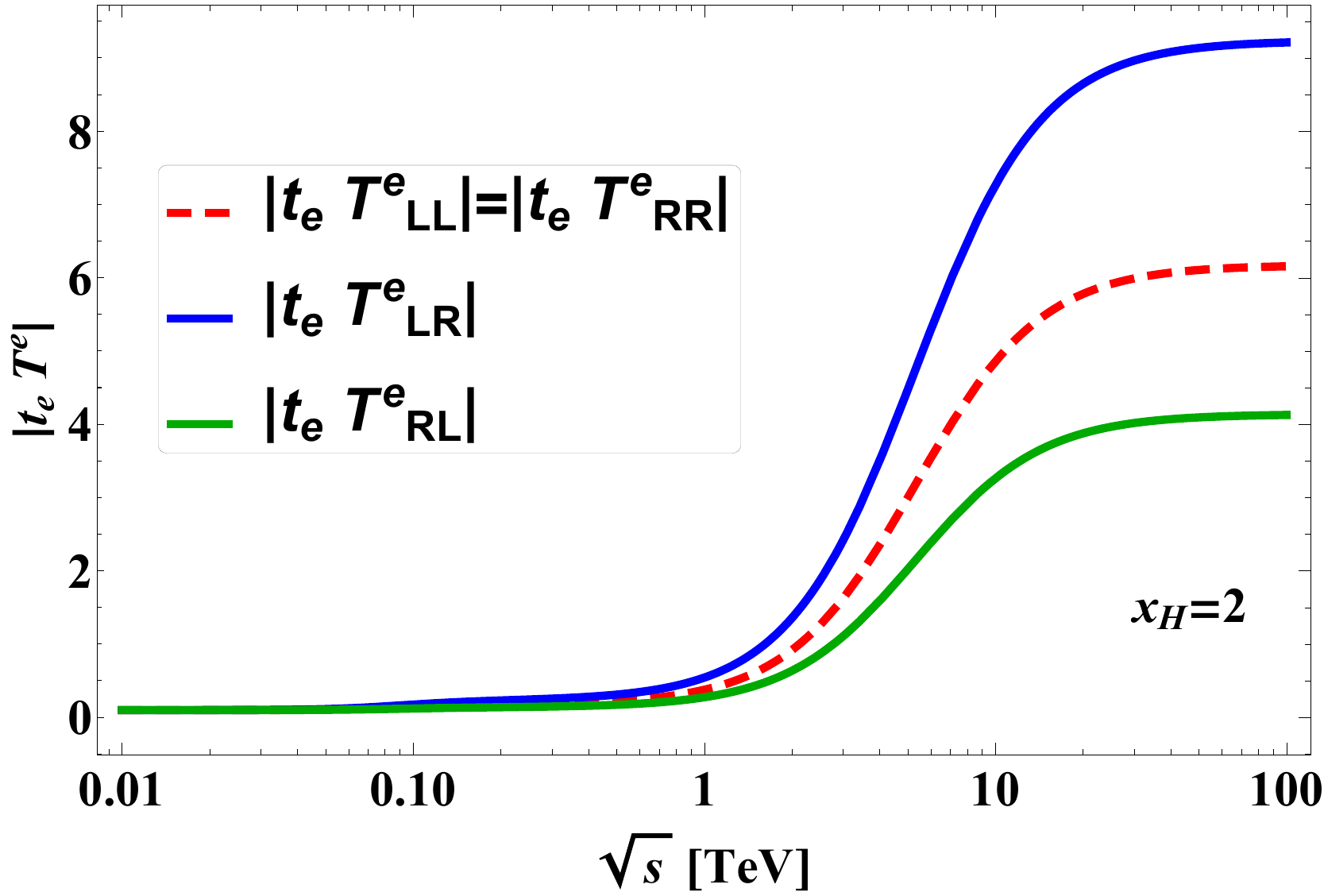}
 \caption{
 $|t_e T^e|$ as a function of $\sqrt{s}$ for $\mu^+ e^- \to \mu^+ e^-$ process with $\cos\theta = 0.5$ for $M_{Z^\prime}=7.5$ TeV and different $x_H$.
 } 
\label{fig-tTe}
\end{center}
\end{figure} 
%%%%%%%%%%%%%%%%%%%%%%
We present the quantities $|t_e T^e_{ij}|$ with $(i,j)=(L,R)$ in Fig.~\ref{fig-tTe} for different $x_H$. The contributions from $T_{RR}^e$ and $T_{LL}^{e}$ coincide with each other because $g_L^{V_\mu}=g_L^{V_e}$ and $g_R^{V_e}= g_R^{V_\mu}$ because our model set-up is generation independent. On the other hand contributions from $T_{RL}^e$ and $T_{LR}^e$ are different because they depend on $g_{L}^{V_\mu} g_{L}^{V_e}$ and $g_{R}^{V_\mu} g_{R}^{V_e}$ respectively. However, these contributions depend on $x_H$. We find that left handed lepton doublets do not interact with $Z^\prime$ when $x_H=-2$ whereas right handed leptons do not interact with with $Z^\prime$ when $x_H=-1$. 
In the case of $x_H=0$, the left and right handed leptons interact equally. Due to this nature the pattern of contribution is different from other charges. In the case of $x_H=1$ and $2$, interactions of all left and right handed leptons with $Z^\prime$ are different manifesting chiral scenarios. Therefore the pattern of contributions from $x_H=1$ and $x_H=2$ are the same, however, different in magnitude. The quantities $|t_e T_{ij}|$ with $(i,j)=(L, R)$ involve effects from the $t-$channel process involving the individual contributions and corresponding interference of $\{\gamma, Z, Z^\prime\}$ mediated interactions which grow with $\sqrt{s}$ with $\cos\theta=0.5$ and $M_{Z^\prime}=7.5$ TeV and after $\sqrt{s}=10$ TeV, the contributions become flat. 

Using the quantities given in Eq.~\ref{Te}, we can obtain the differential scattering cross sections with polarized initial states as the followings,  
\begin{eqnarray}
\frac{d \sigma_{\mu_R e_R}}{d \cos \theta} 
= \frac{u_e^2}{8\pi s_e}\left| T_{RR}^e \right|^2, \, \, \,
\frac{d \sigma_{\mu_L e_L}}{d \cos \theta} 
= \frac{u_e^2}{8\pi s_e}\left| T_{LL}^e \right|^2, \,\,\,
\frac{d \sigma_{\mu_R e_L}}{d \cos \theta} 
= \frac{s_e}{8\pi}\left| T_{RL}^e \right|^2, \, \, \,
\frac{d \sigma_{\mu_L e_R}}{d \cos \theta} 
= \frac{s_e}{8\pi}\left| T_{LR}^e \right|^2. 
\end{eqnarray}
When initial muon and electron are longitudinally polarized, the differential scattering cross section is given by 
\begin{eqnarray}
\frac{d\sigma^e}{d\cos \theta}(P_\mu, P_e) &=&
 \frac14 \left(
(1+P_\mu)(1+P_e)\frac{d \sigma_{\mu_R e_R}}{d \cos \theta}
+
(1-P_\mu)(1-P_e)\frac{d \sigma_{\mu_L e_L}}{d \cos \theta}
\right. \nonumber \\ 
&& \left.
+
(1+P_\mu)(1-P_e)\frac{d \sigma_{\mu_R e_L}}{d \cos \theta}
+
(1-P_\mu)(1+P_e)\frac{d \sigma_{\mu_L e_R}}{d \cos \theta}
\right), 
\label{X-sec1}
\end{eqnarray}
where $P_\mu$ and $P_e$ are the polarization of the muon and electron beams. 
$P_i = +1$ denotes purely right-handed initial particles and $P_i = -1$ denotes purely left-handed initial particles. The deviation of the differential cross section of $U(1)_X$ case from the SM scenario is given by 
\begin{eqnarray}
\Delta_{d\sigma^e}
\equiv
\frac{\left[\frac{d\sigma^e}{d\cos \theta}(0, 0)\right]_{U(1)_X}}{\left[\frac{d\sigma^e}{d\cos \theta}(0, 0)\right]_{SM}}-1.
\end{eqnarray}
%%%%%%%%%%%%%%%%%%%%%
\begin{figure}[h]
\begin{center}
\includegraphics[width=0.6\textwidth,angle=0]{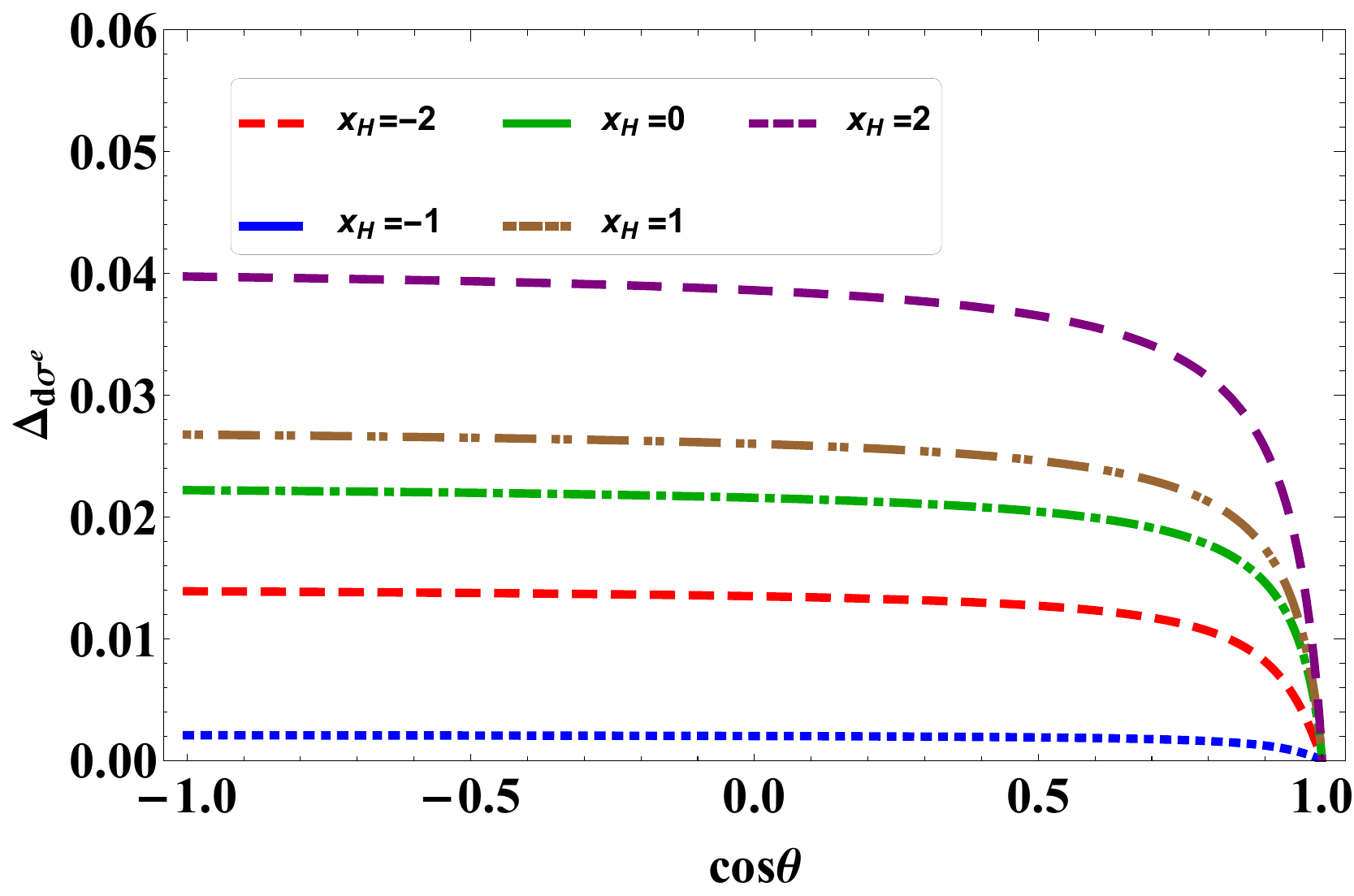}
 \caption{Deviation of total cross sections for $\mu^+ e^- \to \mu^+ e^-$ process from the SM as a function of $\cos\theta$ for $\sqrt{s}=346$ GeV and $M_{Z^\prime}=7.5$ TeV with 1 ab$^{-1}$ using different $x_H$.
 } 
\label{fig-dsigmae}
\end{center}
\end{figure} 
%%%%%%%%%%%%%%%%%%%%%%
Fig.~\ref{fig-dsigmae} shows the deviation of the differential cross section from the SM showing the difference of the charges as a function of $\cos\theta$. We find that the deviation grows with $\cos\theta$ and reaches a saturation when $\cos\theta < 0.2$. We find that the deviation can be maximum for $x_H=2$ reaching at $4\%$ whereas that is minimum for $x_H=-1$ and the value is $0.2\%$ where right handed leptons have no interaction with $Z^\prime$. We also mention that the deviation reaches at $1.6\%$ for $x_H=-2$, that for $x_H=0$ is $2.2\%$ and $2.75\%$ for $x_H=1$, respectively where $x_H=-2$ is the case where left handed lepton doublets do not interact withe $Z^\prime$ and $x_H=0$ is the B$-$L case. The deviation from the SM results could reach up to $4\%$ for $x_H=2$. Such deviations carry contributions from the $Z^\prime$ mediated process and its interference with the $Z$ and photon mediated modes. The deviations from the SM falls sharply for $\cos\theta \geq 0.9$ and at the vicinity of $1.0$, the effect from different charges are indistinguishable.

\begin{sidewaysfigure}
\includegraphics[width=0.33\textwidth,angle=0]{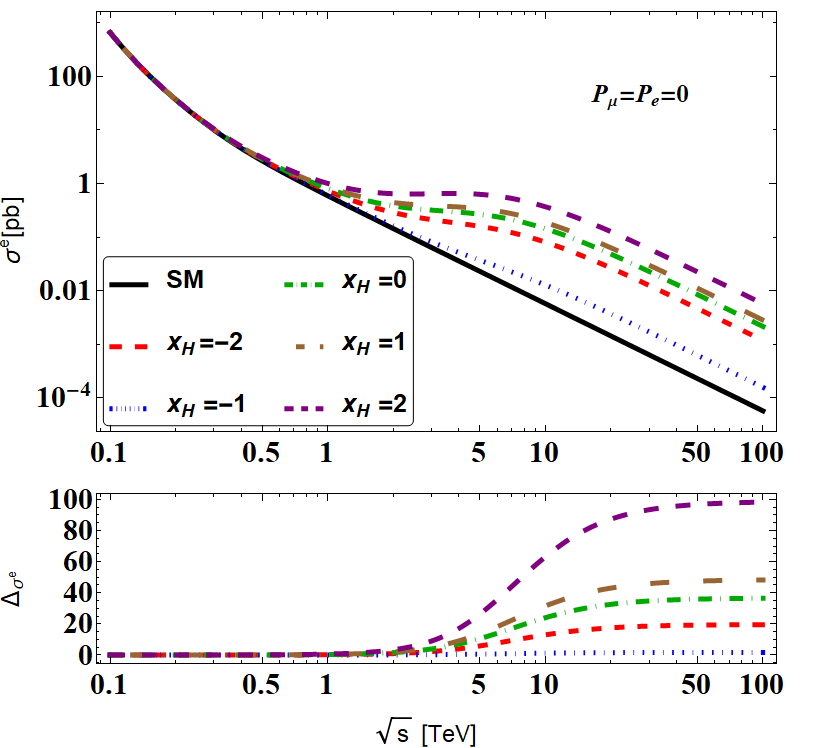}
\includegraphics[width=0.33\textwidth,angle=0]{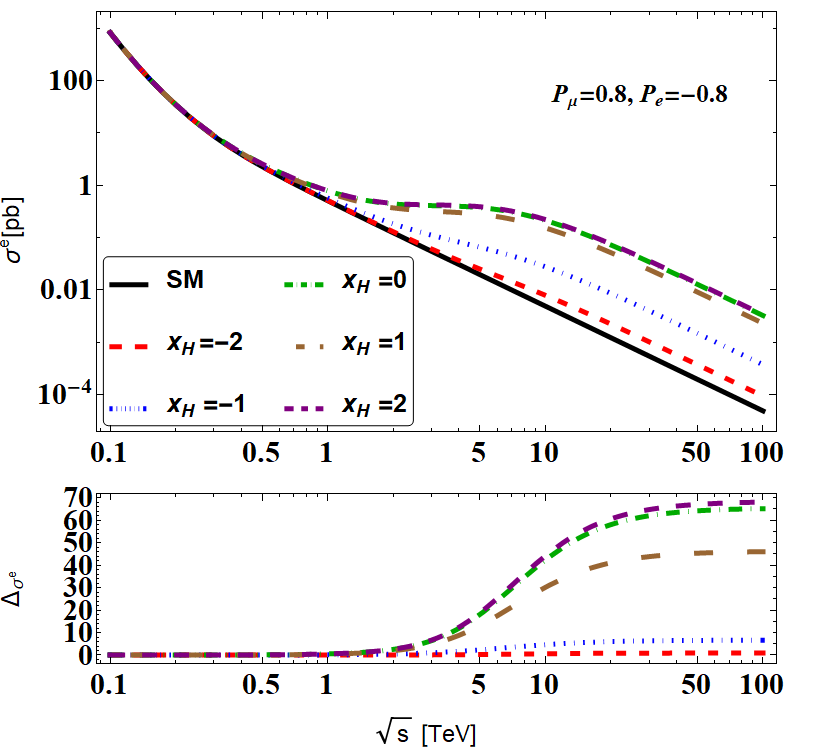}
\includegraphics[width=0.33\textwidth,angle=0]{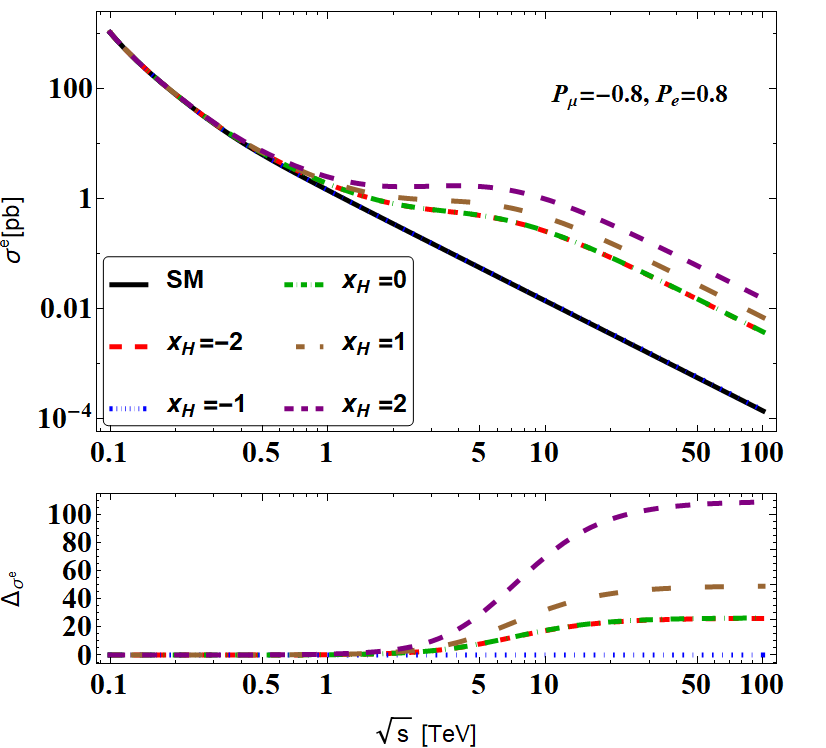}\\
\includegraphics[width=0.32\textwidth,angle=0]{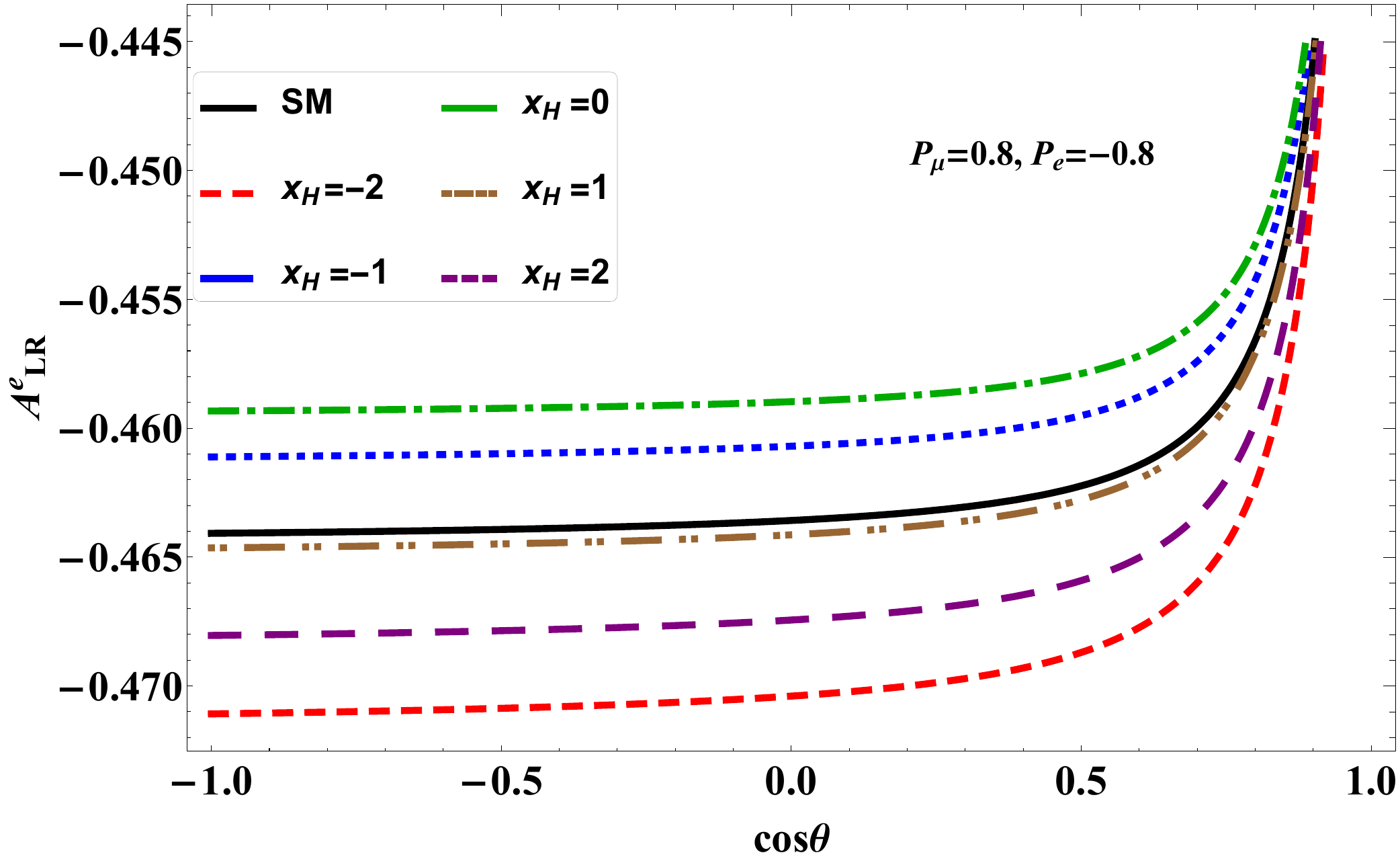}
\includegraphics[width=0.32\textwidth,angle=0]{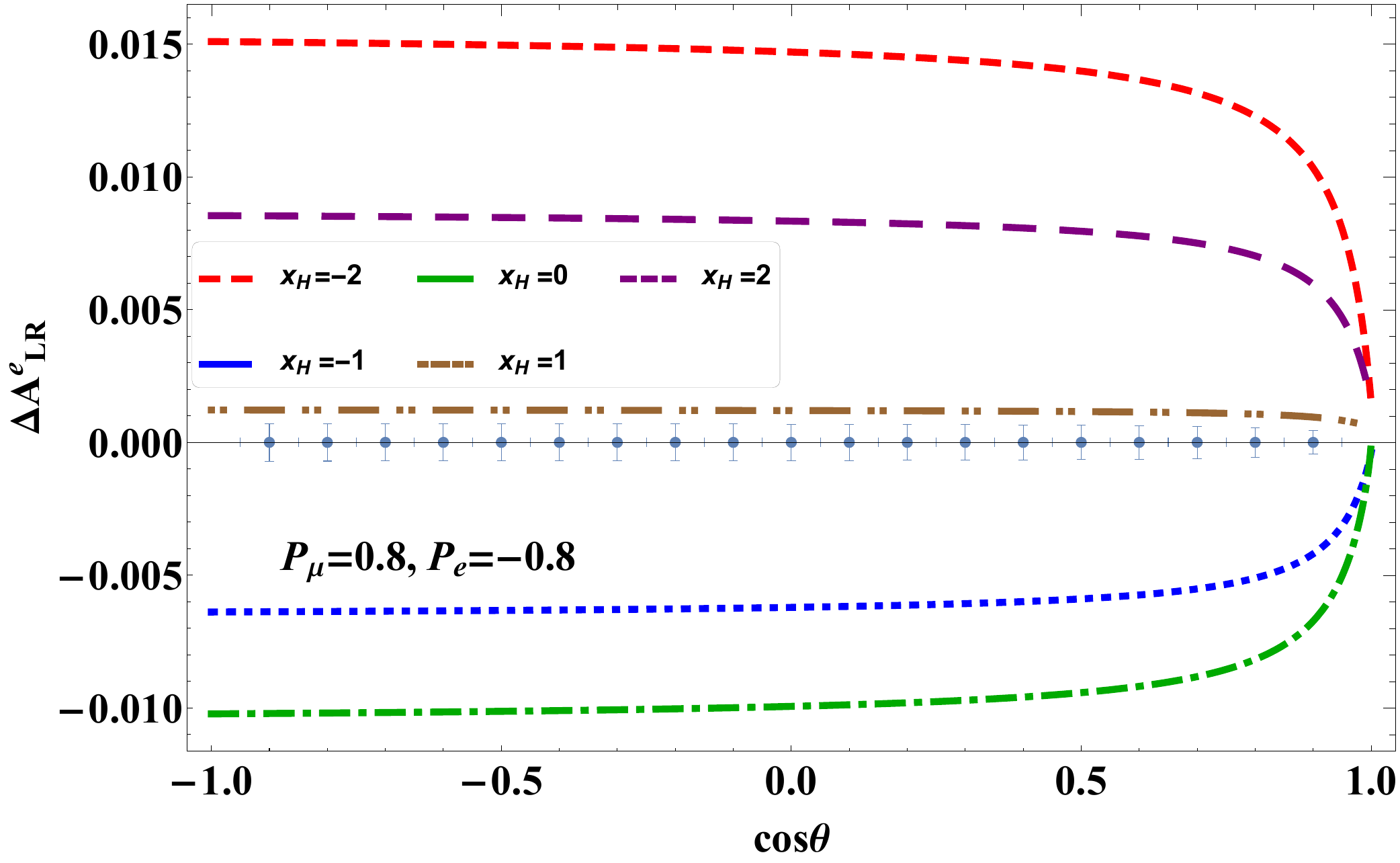}
\includegraphics[width=0.32\textwidth,angle=0]{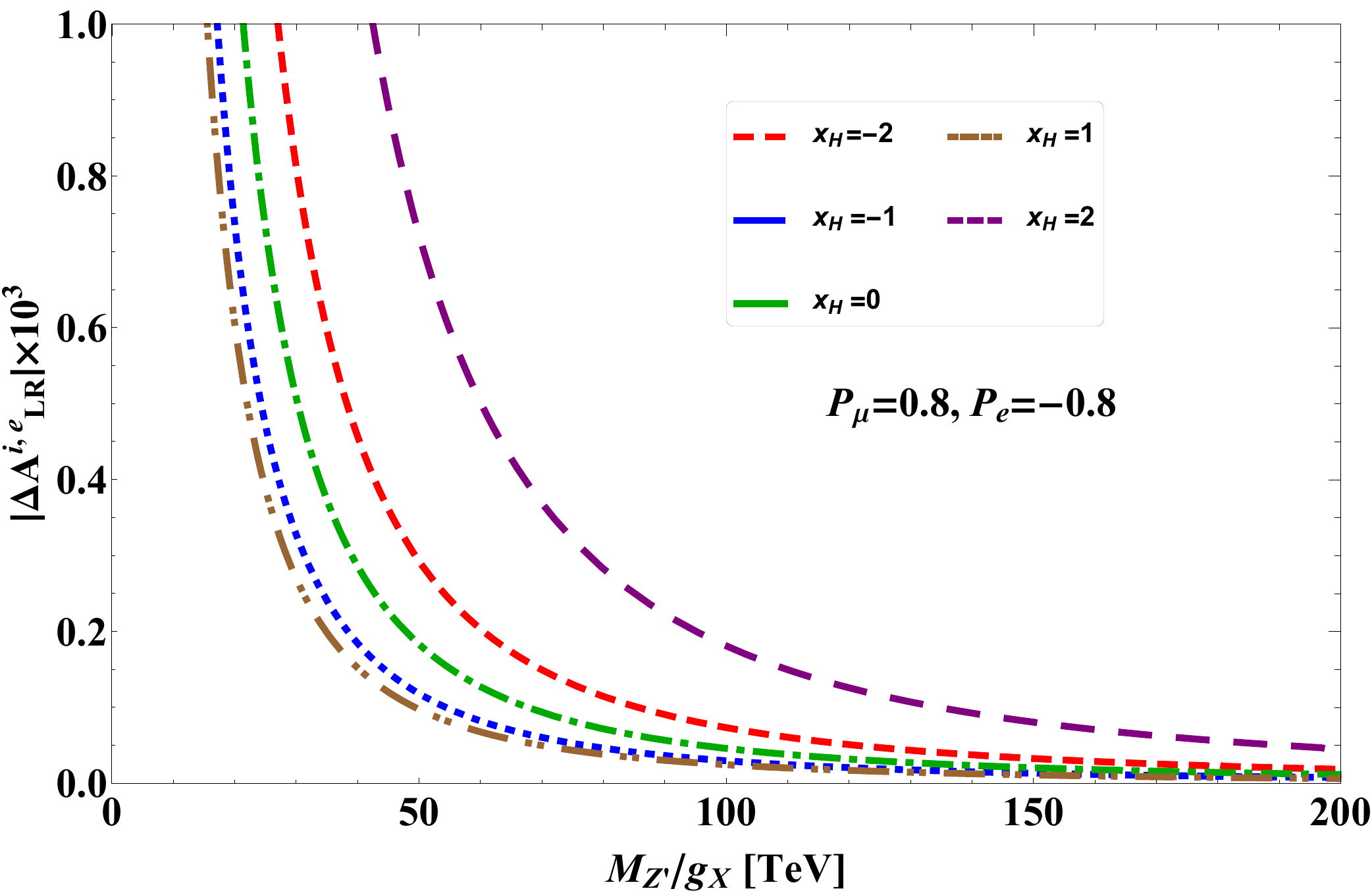}
 \caption{The total cross section for $\mu^+ e^- \to \mu^+ e^-$ process and the corresponding deviation are shown in the upper panel from the SM for different polarization choices as a function of $\sqrt{s}$ for $M_{Z^\prime}= 7.5$ TeV and different $x_H$ for $P_{\mu}=P_{e}=0$, $P_{\mu}=0.8, P_{e}=-0.8$ and $P_{\mu}=-0.8, P_{e}=0.8$ from left to right respectively. In the lower panel we show the differential left-right asymmetry for $\mu^+ e^- \to \mu^+ e^-$ scattering process with respect to $\cos\theta$ in the lower left panel whereas its deviation with respect to $\cos\theta$ are given in the lower middle panel for different $x_H$. Deviation in integrated left-right asymmetry for this process as a function of $M_{Z^\prime}/g_X$ for different values of $x_H$ are shown in the lower right panel. To estimate the differential left-right asymmetry and the corresponding deviations we take $M_{Z^\prime}=7.5$ TeV. In this analysis we use $\sqrt{s}=346$ GeV in appropriate places.} 
\label{fig-sigmae}
 \end{sidewaysfigure}

The total scattering cross section and the deviation of the total scattering cross section under the $U(1)_X$ case from the SM result for the $\mu^+ e^- \to \mu^+ e^-$ process are given by 
\begin{eqnarray}
\sigma^e(P_\mu, P_e)
&\equiv&
\int_{-0.99}^{0.99}
\frac{d\sigma^e}{d\cos \theta}(P_\mu, P_e)
d(\cos\theta), 
\\ 
\Delta_{\sigma^e}
&\equiv&
\frac{\left[\sigma^e(P_\mu, P_e)\right]_{U(1)_X}}{\left[\sigma^e(P_\mu, P_e)\right]_{SM}}-1, 
\end{eqnarray}
respectively. Since the cross section diverges at $\cos\theta \simeq 1$ therefore we integrate the differential scattering cross section within the range $-0.99 \leq \cos\theta \leq 0.99$ for different $x_H$ and SM for different polarization of muon and electron as $P_{\mu}=P_{e}=0$, $P_{\mu}=0.8, P_{e}=-0.8$ and $P_{\mu}=-0.8, P_{e}=0.8$. The corresponding scattering cross  sections for different $x_H$ are shown in the upper panel of Fig.~\ref{fig-sigmae} with the corresponding deviations as a function of $\sqrt{s}$ with $M_{Z^\prime}=7.5$ TeV for different polarization modes mentioned above from left to right. In Eq.~(\ref{X-sec1}), the dominant contribution comes from $\frac{d(\sigma_{\mu_L e_R})}{d(\cos \theta)}$ because $(1-(-0.8))(1+0.8)=3.24$ is larger than $(1-(-0.8))(1-0.8)=(1+(-0.8))(1+0.8)=0.36$ and $(1+(-0.8))(1-0.8)=0.04$ respectively. In addition to that, $T_{LR}$ for $x_H = 0$ and $-2$ are the same because $g_R{V_\mu} g_R^{V_e}=(\pm g_X)^2$ where plus sign corresponds to $x_H=-2$ and minus sign corresponds to $x_H=0$. Therefore the cross sections are similar. The BSM case consists of the contribution from the $Z^\prime$ mediated process, in addition to $Z$ and photon mediated processes including the contributions from corresponding interference terms. The SM result is shown by the black solid line which decreases with the increase in $\sqrt{s}$ due to the absence of the contribution from $Z^\prime$.
%%%%%%%%%%%%%%%%%%%%%
%%%%%%%%%%%%%%%%%%%%%%
The cross sections start deviations from the SM results and attains a maximum at $\sqrt{s}=M_{Z^\prime}=7.5$ TeV and slowly decreases when $\sqrt{s}> M_{Z^\prime}$. The deviations of these cross sections stay within $5\%$ for $\sqrt{s}=346$ GeV and slowly grows with the increase in $\sqrt{s}$ and finally maintains a constant value up to $\sqrt{s}=100$ TeV. For $\sqrt{s} > M_{Z^\prime}$ the deviation varies between $25\%-100\%$ depending on $x_H$ for different polarization of incoming muon and electron. 

The $\mu^+ e^- \to \mu^- e^+$ scattering mediated by $Z^\prime$, $Z$ and photon are forward dominant $t-$channel processes. Therefore we study the left-right asymmetry $(\mathcal{A}_{\rm{LR}}^e)$ of polarized cross sections and the differential left-right asymmetry is given by 
\begin{eqnarray}
\mathcal{A}^e_{\rm{LR}}(P_\mu, P_e, \cos \theta) \equiv 
 \frac{\frac{d\sigma^e}{d\cos\theta}(P_\mu, P_e) - \frac{d\sigma^e}{d\cos\theta}( - P_\mu, - P_e)}{\frac{d\sigma^e}{d\cos\theta}(P_\mu, P_e) + \frac{d\sigma^e}{d\cos\theta}( - P_\mu, - P_e)},  
\end{eqnarray}
and the deviation of the differential left-right asymmetry from the SM is given by 
\begin{eqnarray}
\Delta \mathcal{A}^e_{\rm{LR}}\equiv \frac{[A^e_{LR}]_{U(1)_X}}{[A^e_{LR}]_{SM}} - 1. 
\end{eqnarray}
%%%%%%%%%%%%%%%%%%%%%%%%%%%%%%%%%%
The differential left-right asymmetry and its deviation with respect to $\cos\theta$ are shown in the lower left and middle panels of Fig.~\ref{fig-sigmae} respectively for different $x_H$ with $P_{\mu}=0.8$, $P_{e}=-0.8$. We find that differential left-right asymmetry varies between $-0.471 \leq \mathcal{A}_{\rm LR}\leq -0.445$ for $-0.99 \leq \cos\theta \leq 0.99$ which sharply rises after $\cos\theta \geq 0.5$. We find that for $x_H=-2$ and $-1$ the differential left-right asymmetry may attain the values $-0.471$ and $-0.464$ respectively for $\cos\theta \leq 0.5$. That for the B$-$L case is $-0.461$ for $\cos\theta \leq 0.5$. In the case of $x_H=1$ and $2$ all the left and right handed fermions interact differently with $Z^\prime$, however, they have different values. The SM contribution is shown by the solid black line which deviates from the BSM results when $\cos\theta \leq 0.35$. The deviation in differential left-right asymmetry $(\Delta\mathcal{A}_{\rm{LR}}^e)$ are shown in the top right panel for different $x_H$ with respect to $\cos\theta$. The deviation in differential left-right scattering remains constant for $\cos\theta < 0.5$. The maximum deviation for $x_H=-2$ reaches up to $1.5\%$ whereas that for $x_H=-1$ becomes $0.5\%$, $x_H=0$ becomes $1\%$, $x_H=2$ becomes $0.85\%$ and $x_H=1$ becomes $0.1\%$ respectively. We show the error bar for the deviation in differential left-right asymmetry in the figure from Eq.~\ref{ALR-1} which is small compared to different values of the deviation in left-right asymmetry for different $x_H$.

The integrated left-right asymmetry for the $\mu^+ e^- \to \mu^+ e^-$ process is given by 
\begin{eqnarray}
\mathcal{A}^{i, e}_{\rm LR}
\equiv
 \frac{\int \frac{d\sigma^e}{d\cos\theta}(P_\mu, P_e)d(\cos\theta) - \int \frac{d\sigma^e}{d\cos\theta}( - P_\mu, - P_e)d(\cos\theta)}{\int \frac{d\sigma^e}{d\cos\theta}(P_\mu, P_e)d(\cos\theta) + \int \frac{d\sigma^e}{d\cos\theta}( - P_\mu, - P_e)d(\cos\theta)},  
\end{eqnarray}
and the deviation of the integrated left-right asymmetry from the SM is given by 
\begin{eqnarray}
\Delta \mathcal{A}^{i, e}_{\rm{LR}}\equiv \frac{[A^{i, e}_{LR}]_{U(1)_X}}{[A^{i, e}_{LR}]_{SM}} - 1. 
\label{ALRe-1}
\end{eqnarray}
%%%%%%%%%%%%%%%%%%%%%%%%%%%%%%
The deviation in integrated left-right asymmetry for different $x_H$ with respect to $M_{Z^\prime}/g_X$ are shown in the lower right panel of Fig.~\ref{fig-sigmae} with  $P_{\mu}=0.8$, $P_{e}=-0.8$. The $|\Delta \mathcal{A}_{\rm{LR}}^{i,e}|$ falls sharply for $M_{Z^\prime}/g_X > 50$ TeV for the values of $x_H$ we consider and $M_{Z^\prime}/g_X > 70$ TeV for $x_H=2$, respectively. In this context from we quote the values of $M_{Z^\prime}/g_X$ from LEP-II from Tab.~\ref{tab2} and these are $M_{Z^\prime}/g_X \geq 5.0$ TeV, $2.2$ TeV, $7$ TeV, $11.1$ TeV and $18.0$ TeV for $x_H=-2$, $-1$, $0$, $1$ and $2$, respectively. These bounds are not shown in the lower panel of Fig.~\ref{fig-sigmae} because LEP-II bounds are weaker and they do not cross the curves shown here. We do not add the prospective ILC bounds in the figure, however, they can be found from Tab.~\ref{tab2}. 
%%%%%%%%%%%%%%%%%%%%%%%%%%%%%%%%%%%%%%%
\subsection{$\mu^+ \mu^+$ collider}
%%%%%%%%%%%%%%%%%%%%%%%%%%%%%%%%%%%%%%%
\begin{figure}[h]
\includegraphics[width=0.85\textwidth,angle=0]{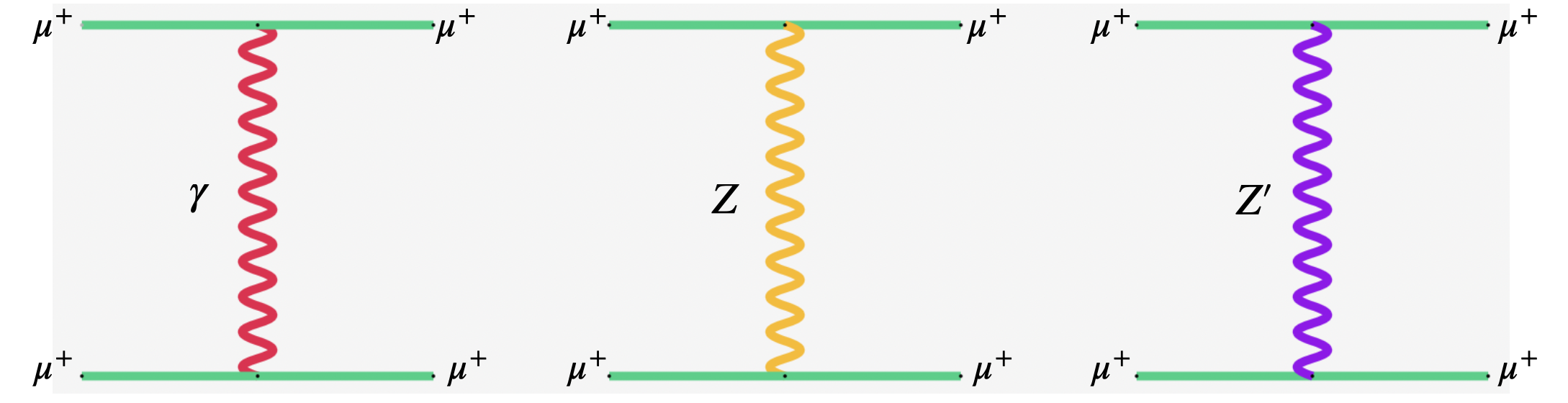}
\includegraphics[width=0.85\textwidth,angle=0]{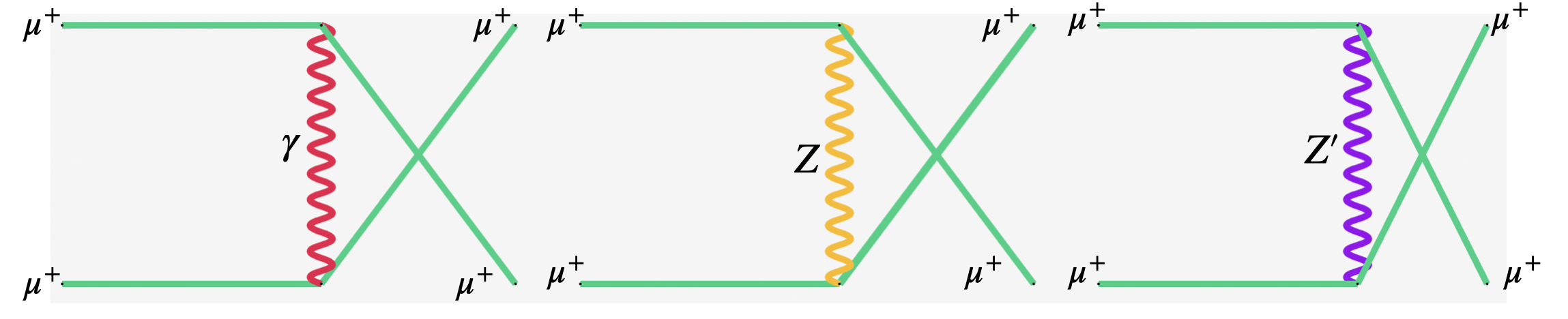}
\caption{Photon, $Z$ and $Z^\prime$ mediated $\mu^+ \mu^+ \to \mu^+ \mu^+$ processes in $t-$channel (upper panel) and $u-$channel (lower panel) at $\mu$TRISTAN experiment with $\sqrt{s}=2$ TeV. } 
\label{fig-mumu}
\end{figure}
%%%%%%%%%%%%%%%%%%%%%%%%%%%%%%%%%%%%
We consider the $\mu^+ \mu^+ \to \mu^+ \mu^+$ scattering in the center-of-mass frame. 
Fig.~\ref{fig-mumu} shows the $\mu^+ \mu^+ \to \mu^+ \mu^+$ scattering processes. 
We define the quantities $p_1$ and $p_2$ as initial momenta of the incoming muons, $k_1$ and $k_2$ are the final muon momenta. The $4-$vector representation of these momenta are
$p_1 = (E, 0, 0, E)$, $p_2 = (E, 0, 0, -E)$, $k_1 = (E, 0, E \sin\theta, E \cos\theta)$ and $k_2 = (E, 0, -E \sin\theta, - E \cos\theta)$, where $E$ is the initial muon energy and $\theta$ is the scattering angle of the muon. There are photon, Z and $Z^\prime$ mediated $t-$ and $u-$channel processes shown in Fig.~\ref{fig-mumu}. Contributions coming from the interference of these processes are also taken into consideration at the time of calculating the cross sections and respective observables. The Mandelstam variables are given by 
\begin{eqnarray}
s_\mu= 4 E^2,\, \, \,t_\mu= - \frac{s_\mu}{2} (1 - \cos\theta),\, \, \,u_\mu= - \frac{s_\mu}{2} (1 + \cos\theta). 
\end{eqnarray}
We fix the energies of incoming muons at $E = $ 1 TeV each and the center-of-mass energy becomes $\sqrt{s_\mu} =$ 2 TeV. We define the following quantities
\begin{eqnarray}
T_{RR}^\mu \equiv 
\sum_{V=\{\gamma, Z, Z'\}} \frac{\left(g_{L}^{V\mu}\right)^2}{t_\mu - M_V^2},\, \, \, 
T_{LL}^\mu \equiv 
\sum_{V=\{\gamma, Z, Z'\}} \frac{\left(g_{R}^{V\mu}\right)^2}{t_\mu - M_V^2}, \nonumber \\
T_{RL}^\mu = T_{LR}^\mu\equiv 
\sum_{V=\{\gamma, Z, Z'\}} \frac{g_{R}^{V\mu} g_{L}^{V\mu}}{t_\mu - M_V^2},~~~~~~~~~ \nonumber \\
U_{RR}^\mu \equiv
\sum_{V=\{\gamma, Z, Z'\}} \frac{\left(g_{L}^{V\mu}\right)^2}{u_\mu - M_V^2}, \,\,\,
U_{LL}^\mu \equiv
\sum_{V=\{\gamma, Z, Z'\}} \frac{\left(g_{R}^{V\mu}\right)^2}{u_\mu - M_V^2}, \,\,\,
\nonumber \\
U_{RL}^\mu = U_{LR}^\mu \equiv
\sum_{V=\{\gamma, Z, Z'\}} \frac{g_{R}^{V\mu} g_{L}^{V\mu}}{u_\mu - M_V^2}.~~~~~~~~~~ 
\end{eqnarray}
%%%%%%%%%%%%%%%%%%%%%
\begin{figure}[h]
\begin{center}
\includegraphics[width=0.497\textwidth,angle=0]{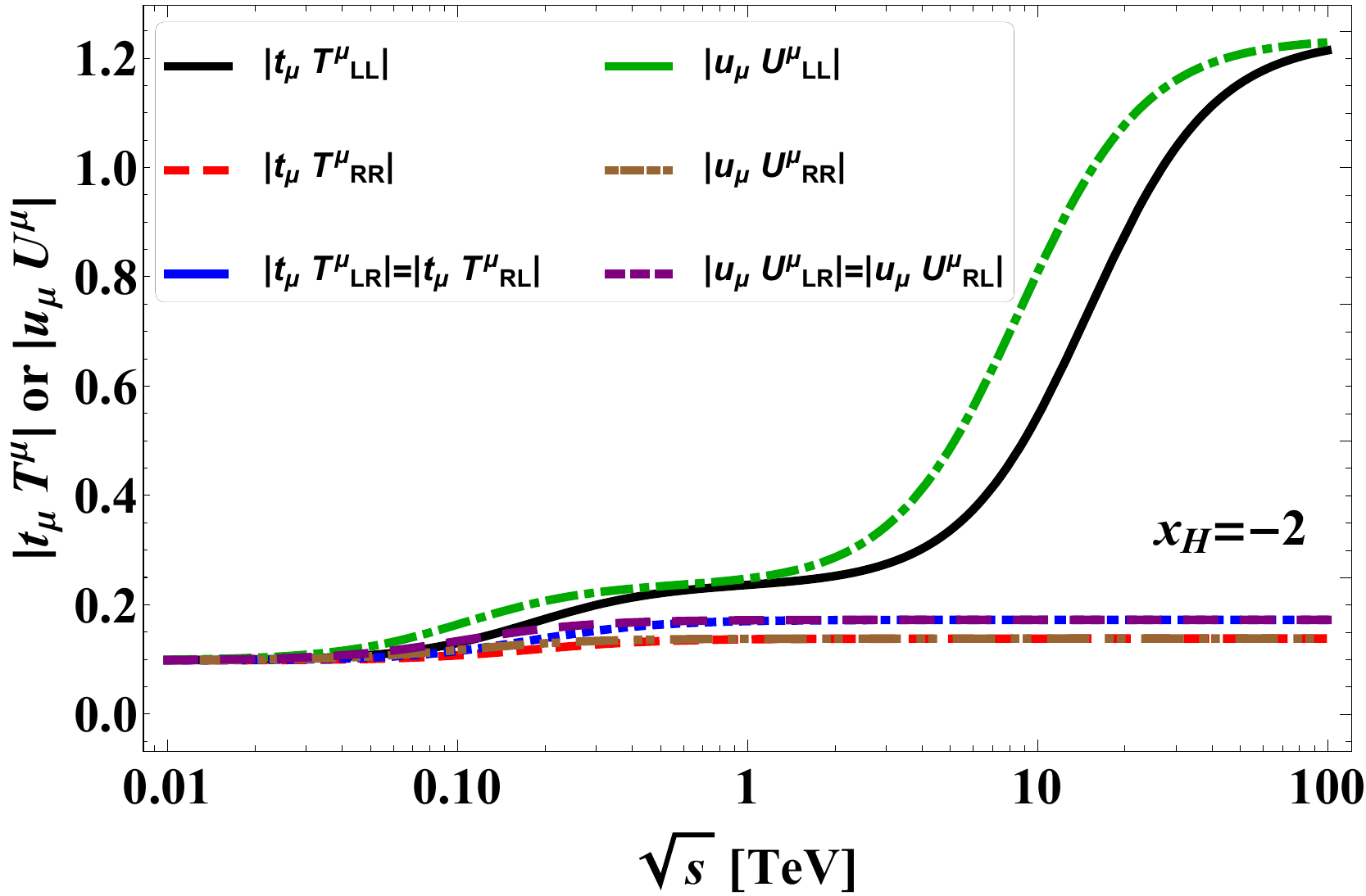}
\includegraphics[width=0.497\textwidth,angle=0]{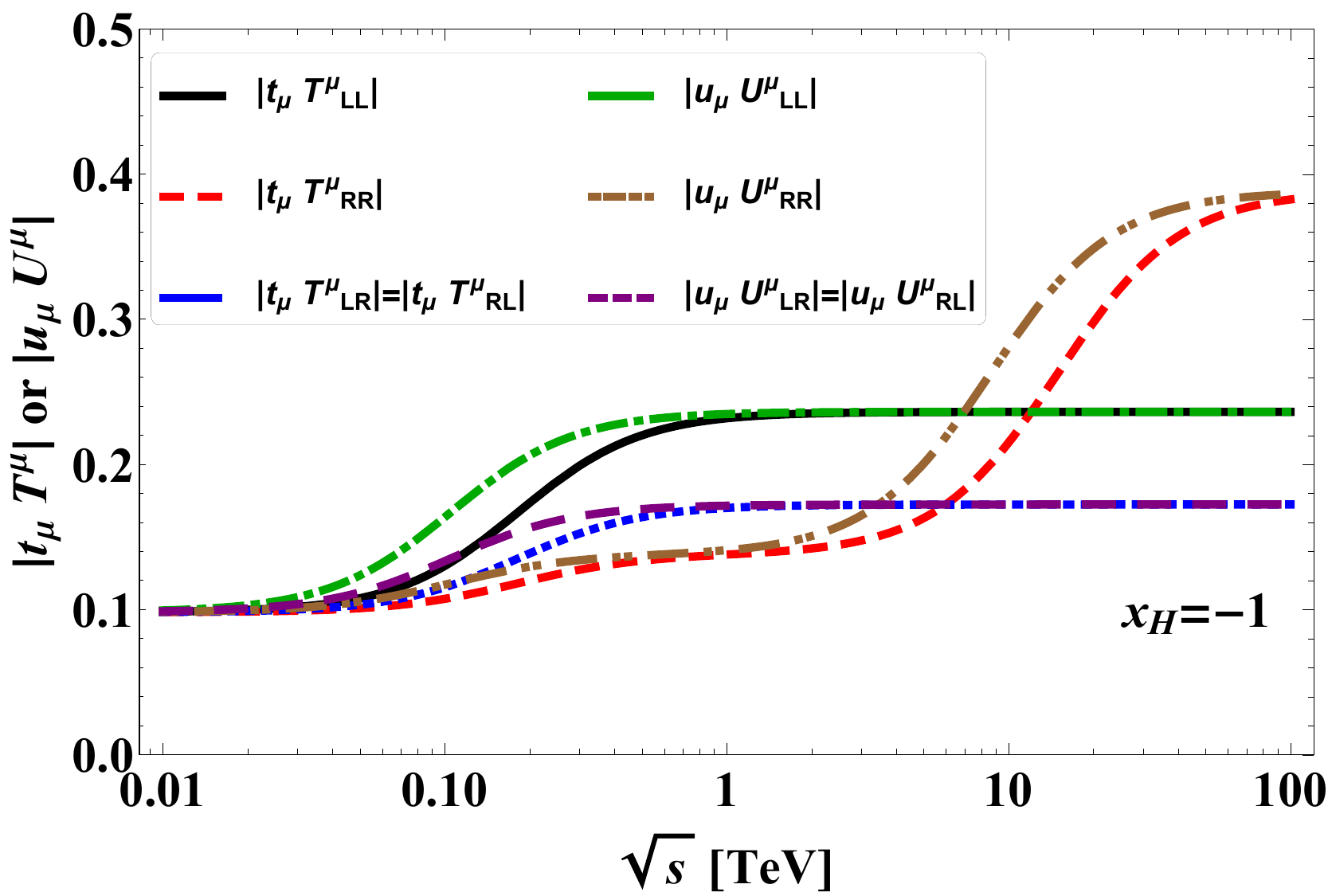}\\
\includegraphics[width=0.497\textwidth,angle=0]{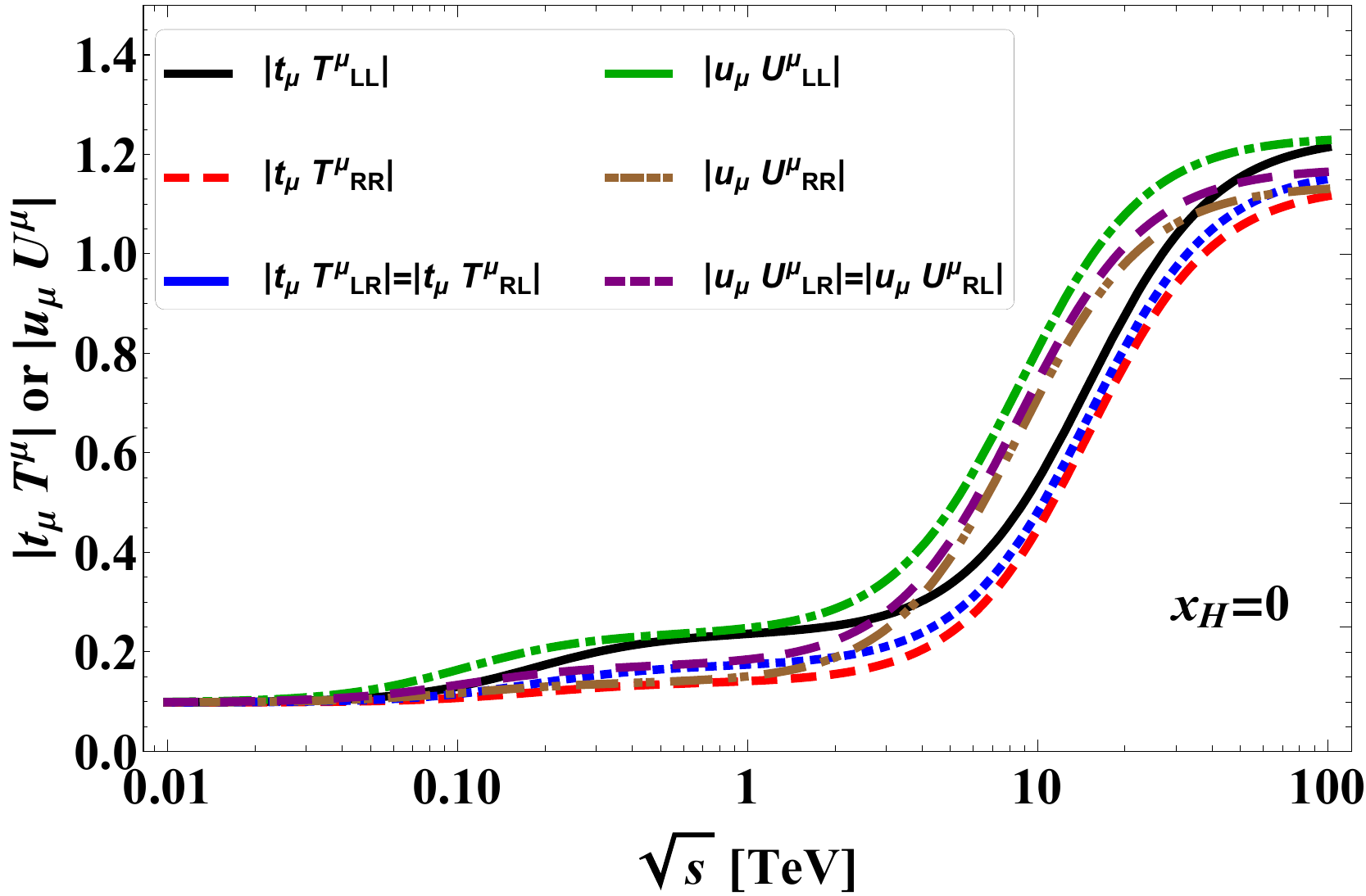}\\
\includegraphics[width=0.497\textwidth,angle=0]{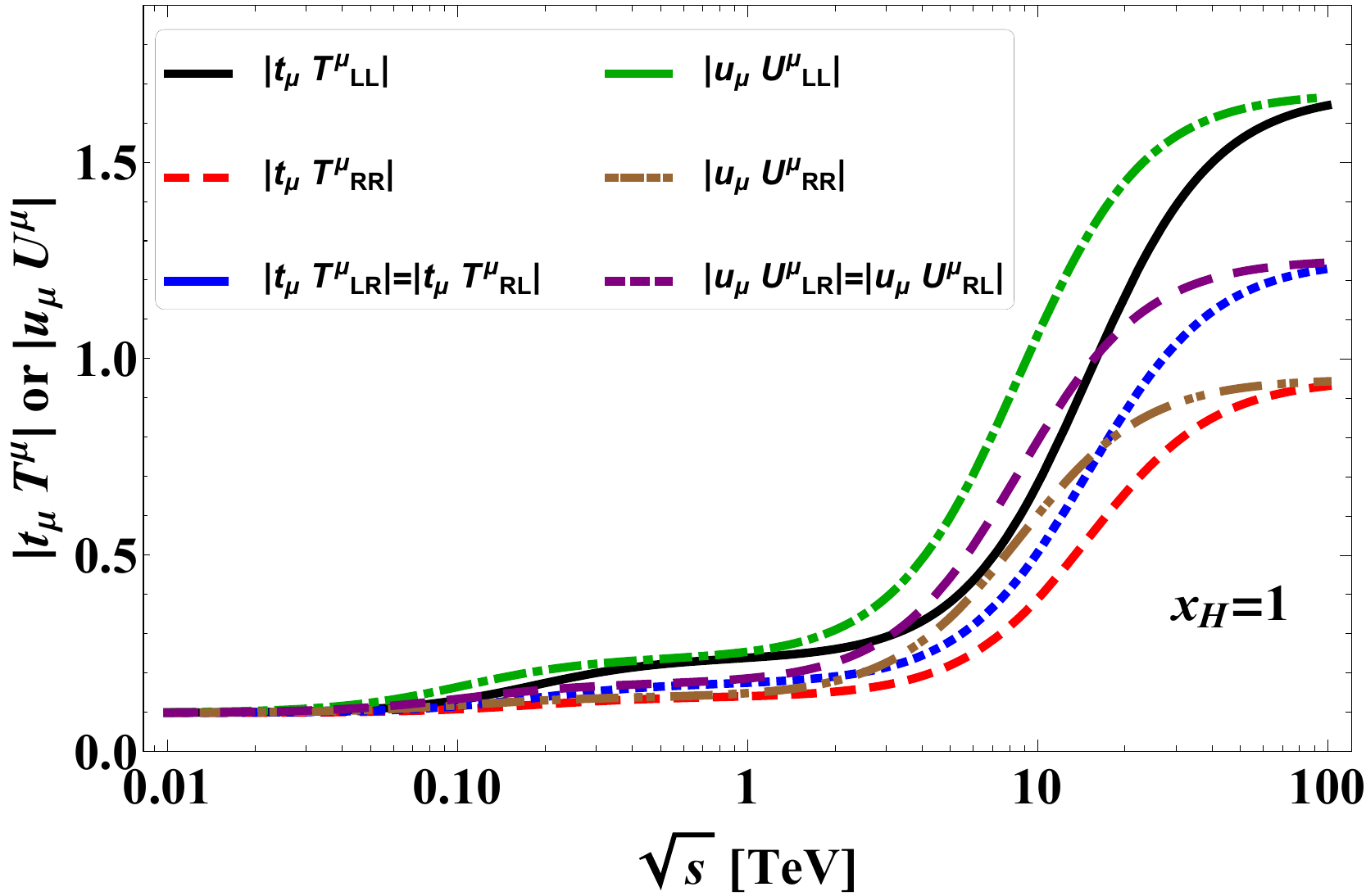}
\includegraphics[width=0.497\textwidth,angle=0]{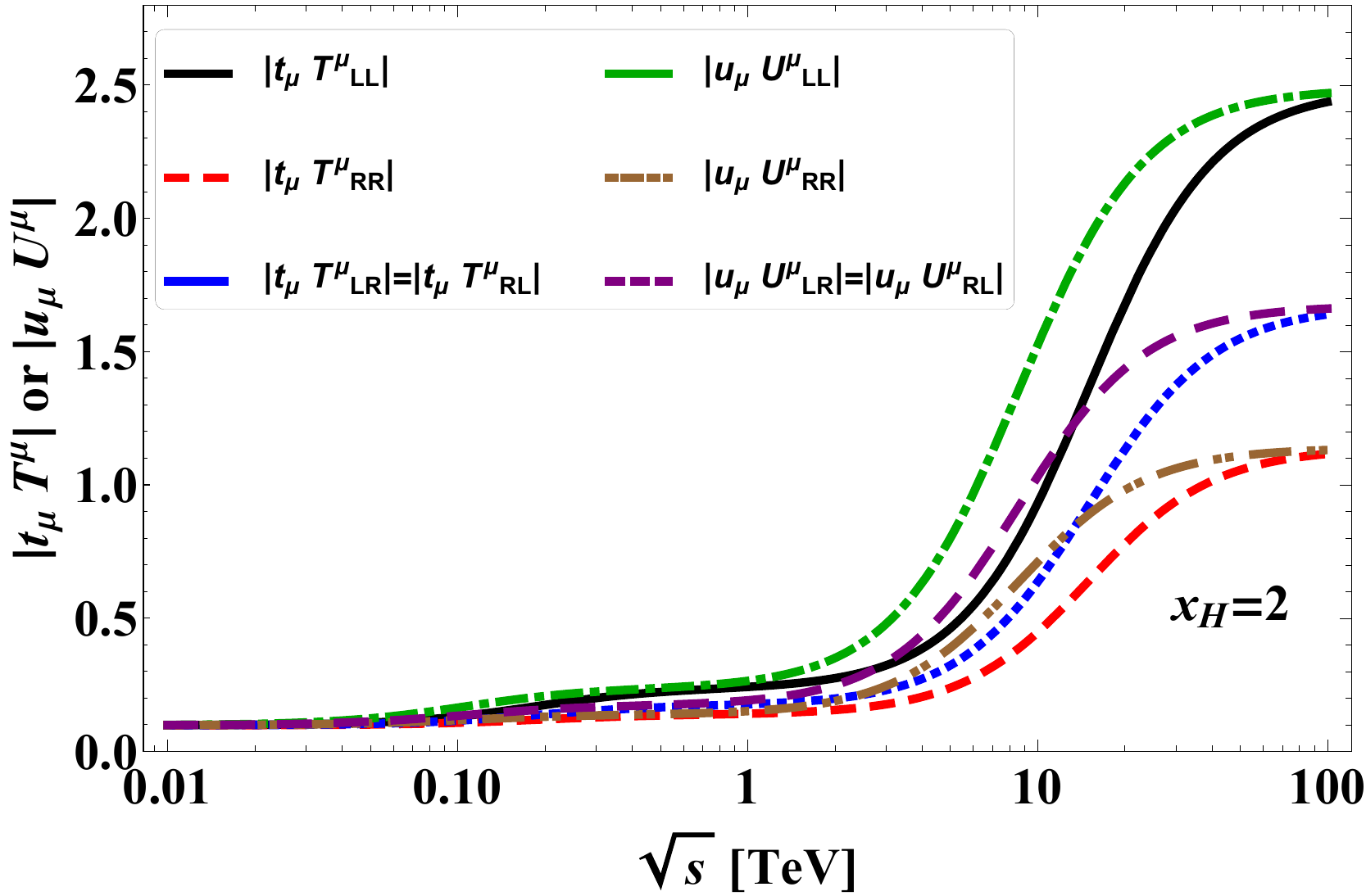}
 \caption{
 $|t_\mu T^\mu|$ and $|u_\mu U^\mu|$ as a function of $\sqrt{s}$ for $\mu^+ \mu^+ \to \mu^+ \mu^+$ scattering process with $\cos\theta = 0.5$ and $M_{Z^\prime}=7.5$ TeV and different $x_H$.
 } 
\label{fig-tTmu}
\end{center}
\end{figure} 
%%%%%%%%%%%%%%%%%%%%%%
We present the quantities $|t_\mu T^\mu|$ and $|u_\mu U^\mu|$ in Fig.~\ref{fig-tTmu}. We find that the effect of the propagators from the $t-$ and $u-$channel processes are same what we observed in $\mu^+e^- \to \mu^+ e^-$ process where we also had $t-$channel mediated processes only. Depending on the $U(1)_X$ charges contributions from the respective propagators increases with increase in $\sqrt{s}$ showing the effects of $t-$ and $u-$channel processes. 

Considering the initial muons are longitudinally polarized, the differential scattering cross sections with polarized initial states are given by 
\begin{eqnarray}
\frac{d \sigma_{\mu_R \mu_R}}{d \cos \theta} 
= \frac{s_\mu}{8\pi}\left| T_{RR}^\mu + U_{RR}^\mu \right|^2, \, \, \,
\frac{d \sigma_{\mu_L \mu_L}}{d \cos \theta} 
= \frac{s_\mu}{8\pi}\left| T_{LL}^\mu + U_{LL}^\mu \right|^2, \, \, \,
\frac{d \sigma_{\mu_R \mu_L}}{d \cos \theta} 
=
\frac{d \sigma_{\mu_L \mu_R}}{d \cos \theta} 
&=& 
\frac{1}{8\pi s_\mu}
\left(
u_\mu^2 \left| T_{RL}^\mu \right|^2 
+ t_\mu^2 \left| U_{RL}^\mu \right|^2\right).~~~~~~
\end{eqnarray}
Hence the differential scattering cross section for the $\mu^+\mu^+ \to \mu^+ \mu^+$ process is given by 
\begin{eqnarray}
\frac{d\sigma^\mu}{d\cos \theta}(P_1, P_2) &=&
 \frac14 \left(
(1+P_1)(1+P_2)\frac{d \sigma_{\mu_R \mu_R}}{d \cos \theta}
+
(1-P_1)(1-P_2)\frac{d \sigma_{\mu_L \mu_L}}{d \cos \theta}
+
2(1 - P_1 P_2)\frac{d \sigma_{\mu_R \mu_L}}{d \cos \theta}\right), 
\end{eqnarray}
where $P_1$ and $P_2$ are the corresponding polarizations of the muon beams under consideration. Here $P_i = +1$ denotes purely right-handed initial particles and $P_i = -1$ denotes purely left-handed initial particles, respectively. Hence we calculate the deviation in differential scattering cross section due the BSM effect from the SM contribution as
\begin{eqnarray}
\Delta_{d\sigma^\mu}
\equiv
\frac{\left[\frac{d\sigma^\mu}{d\cos \theta}(0, 0)\right]_{U(1)_X}}{\left[\frac{d\sigma^\mu}{d\cos \theta}(0, 0)\right]_{SM}}-1.
\end{eqnarray}
%%%%%%%%%%%%%%%%%%%%%
\begin{figure}[h]
\begin{center}
\includegraphics[width=0.497\textwidth,angle=0]{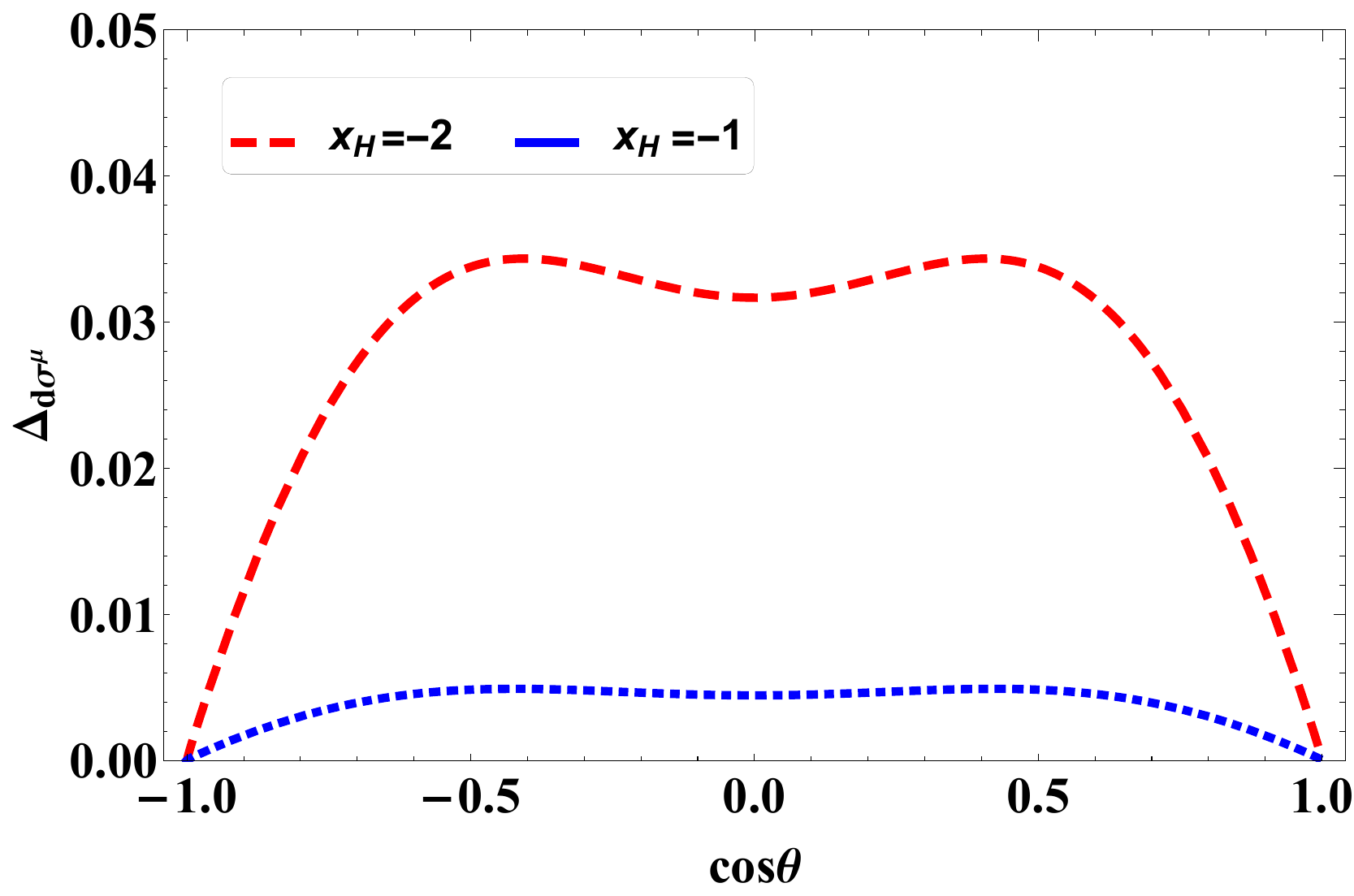}
\includegraphics[width=0.497\textwidth,angle=0]{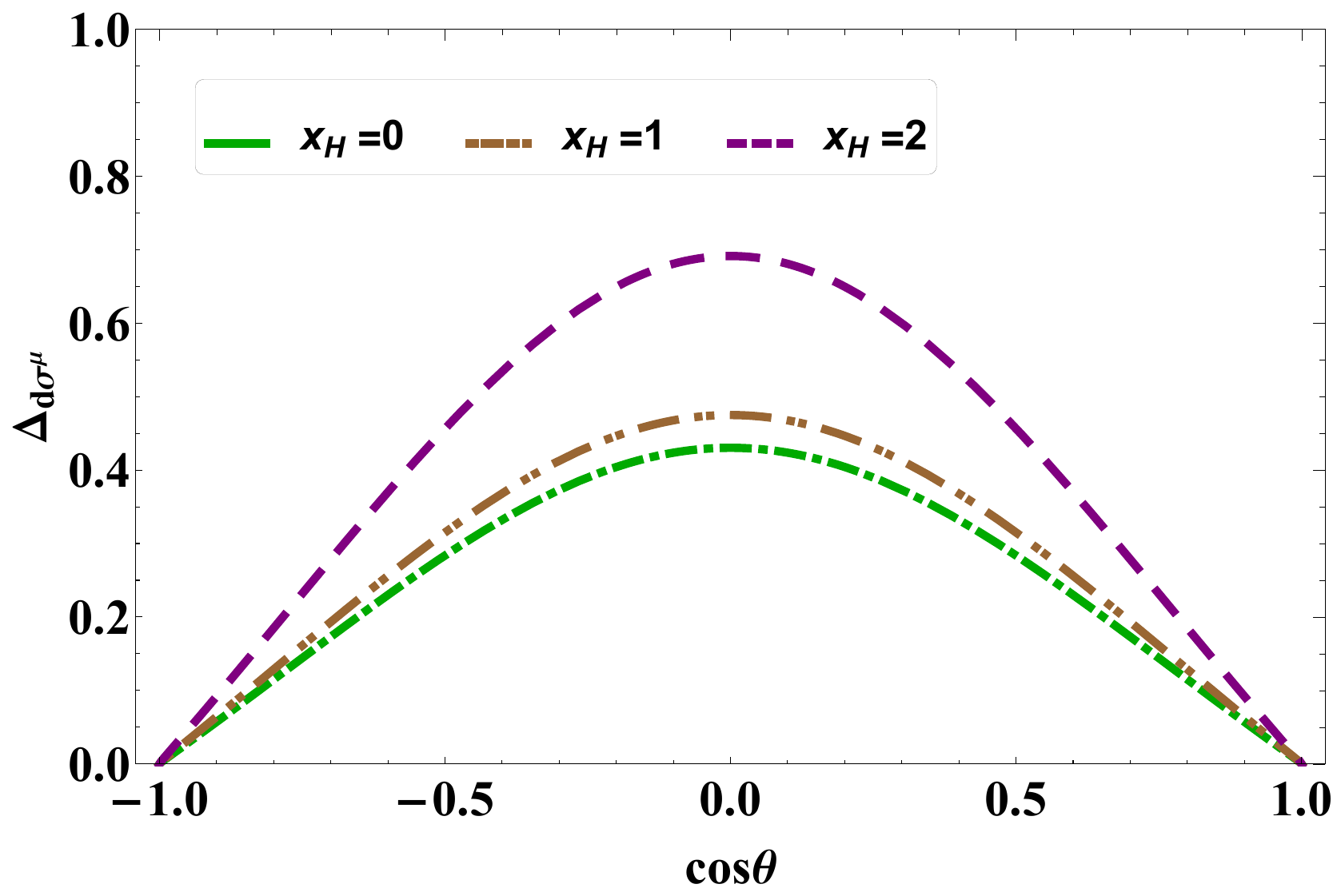}
 \caption{Deviation of total scattering cross sections for $\mu^+ \mu^+ \to \mu^+ \mu^+$ process due to the $Z^\prime$ induced processes from the SM results as a function of $\cos\theta$ at 100 fb$^{-1}$ luminosity for different $x_H$.
 } 
\label{fig-dsigmamu}
\end{center}
\end{figure} 
%%%%%%%%%%%%%%%%%%%%%%
Fig.~\ref{fig-dsigmamu} shows the deviation of the BSM differential scattering cross section due to the effect of $Z^\prime$ under the $U(1)_X$ scenario from the SM. We notice that the deviations for $x_H=-2$ and $-1$ are below $3.6\%$ and $0.5\%$ depending on $\cos\theta$, respectively showing in the left panel where $x_H=-2$ shows the case where left handed lepton doublet has no interaction with $Z^\prime$ whereas for $x_H=-1$, right handed lepton has no interaction with $Z^\prime$. The maximum deviation reaches up to $75\%$ near $\cos\theta\simeq0$ which increases with $x_H$ from zero to 2. All the charges show the effect of $t-$ and $u-$channel processes involving the interference effects.   

The total scattering cross section and the deviation of the total scattering cross section due to the effect of $Z^\prime$ from the SM scenario are given by 
\begin{eqnarray}
\sigma^\mu(P_1, P_2)
&\equiv&
\int_{0}^{0.99}
\frac{d\sigma^\mu}{d\cos \theta}(P_1, P_2)
d(\cos\theta), 
\\ 
\Delta_{\sigma_\mu}
&\equiv&
\frac{\left[\sigma^\mu(P_1, P_2)\right]_{U(1)_X}}{\left[\sigma^\mu(P_1, P_2)\right]_{SM}}-1,
\end{eqnarray}
respectively.
%%%%%%%%%%%%%%%%%%%%%%%%%%%%%%%%%%%%%%
\begin{sidewaysfigure}
\includegraphics[width=0.33\textwidth,angle=0]{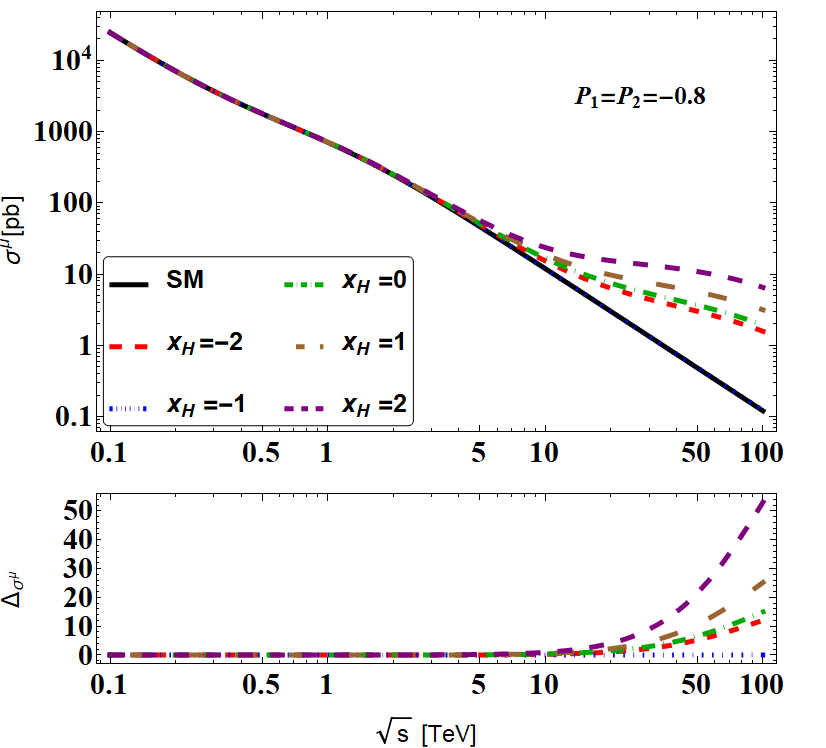}
\includegraphics[width=0.33\textwidth,angle=0]{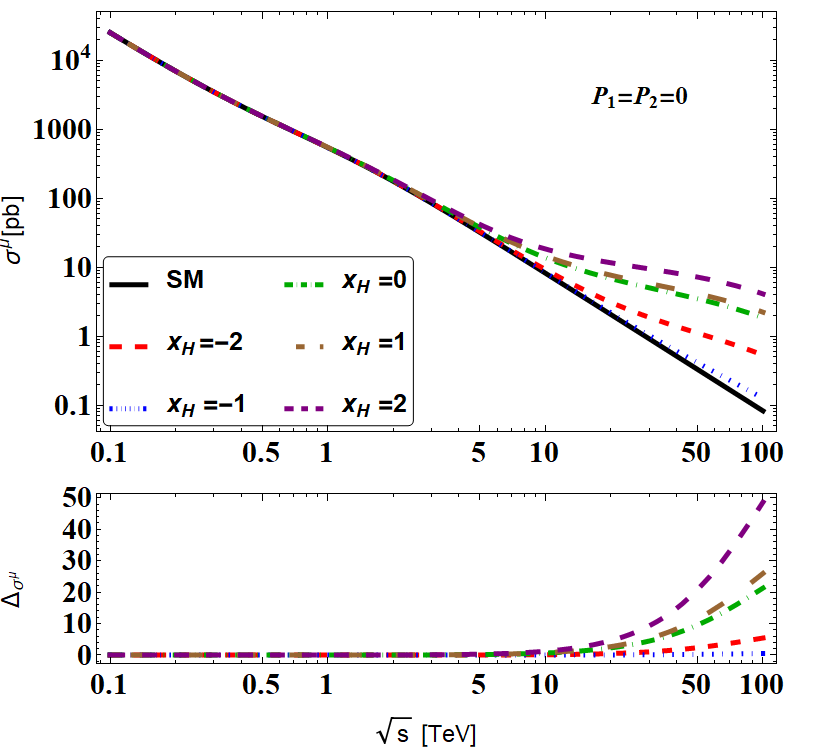}
\includegraphics[width=0.33\textwidth,angle=0]{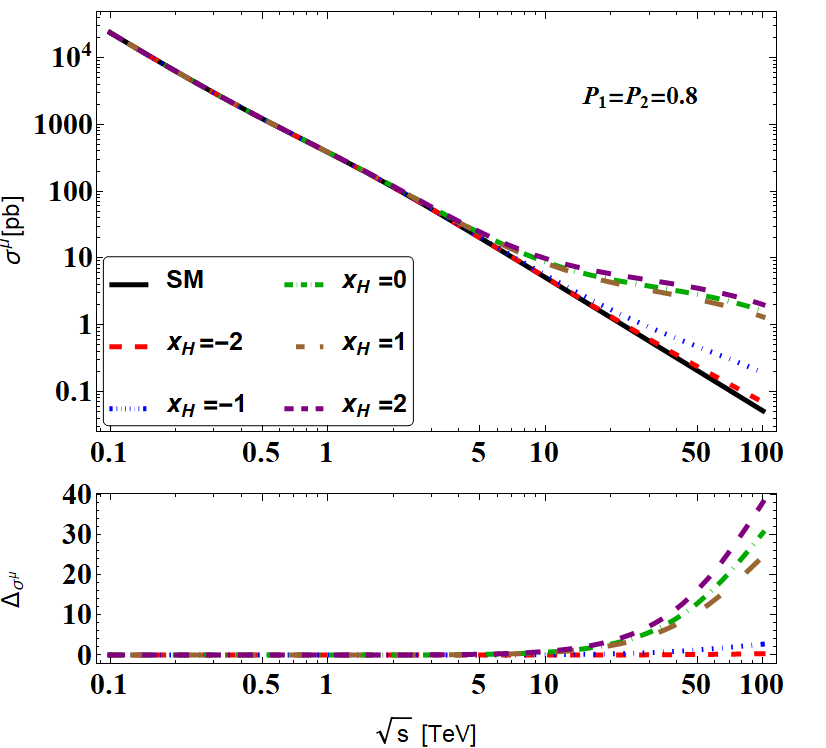}\\
\includegraphics[width=0.32\textwidth,angle=0]{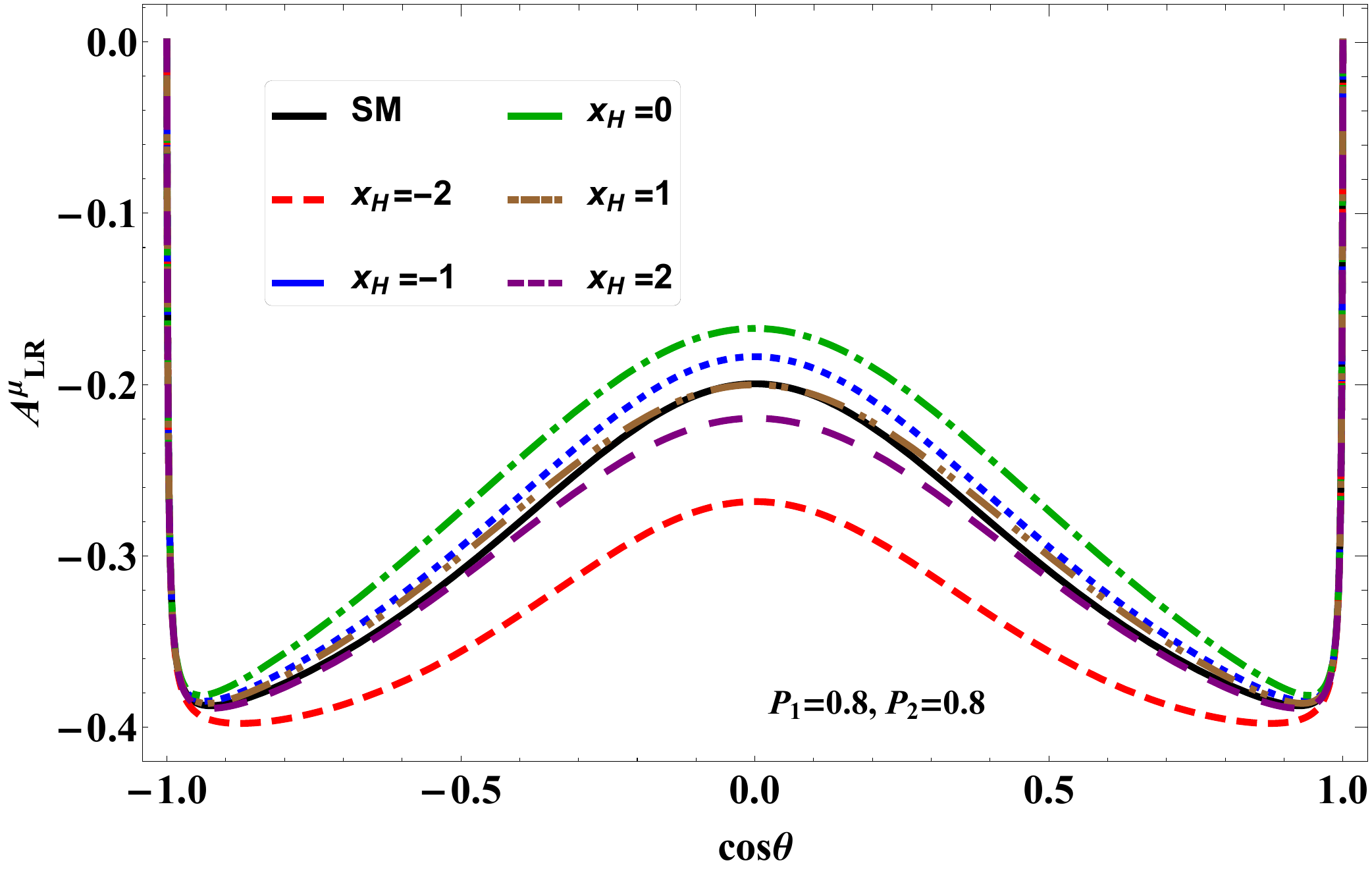}
\includegraphics[width=0.32\textwidth,angle=0]{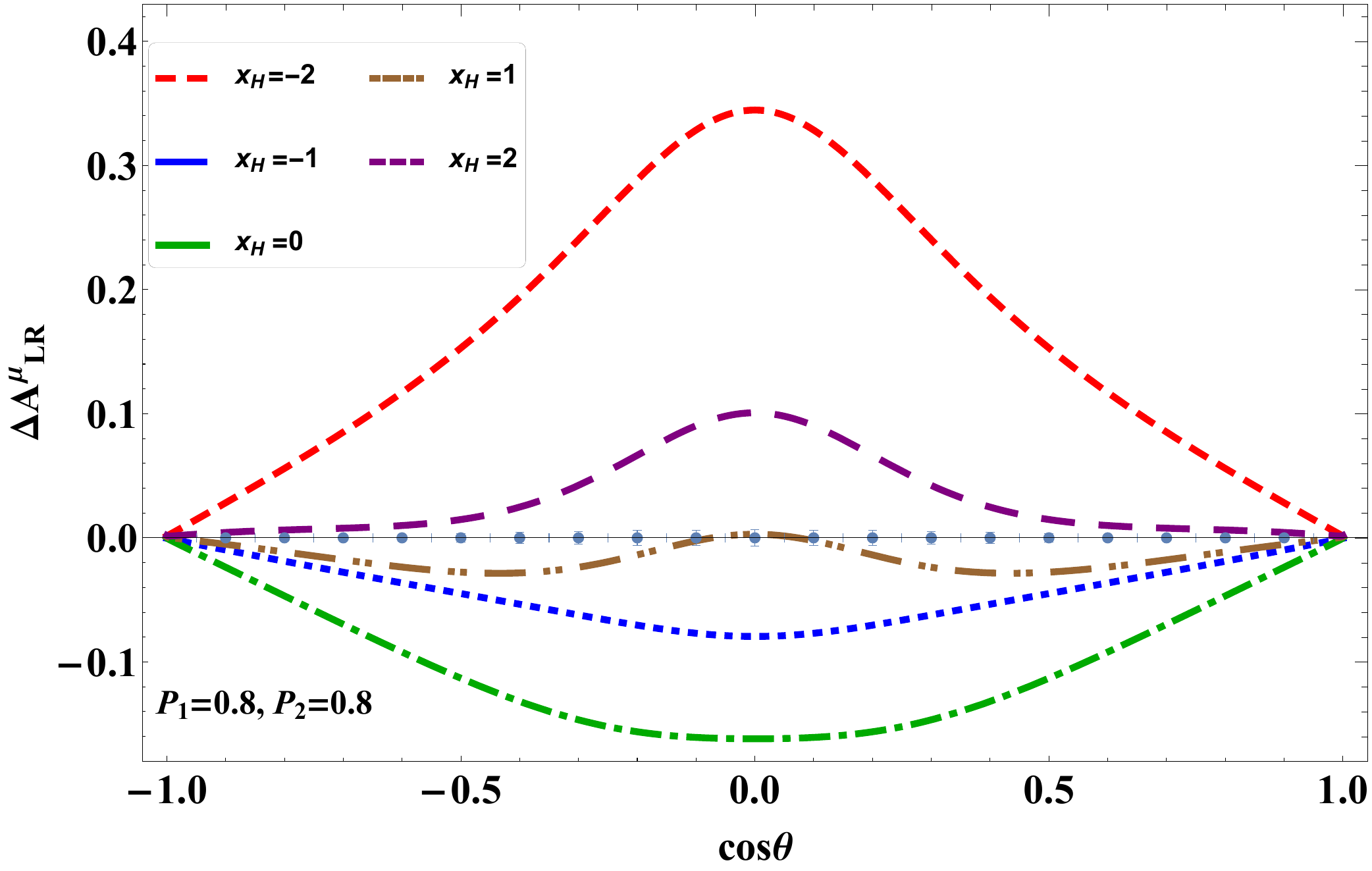}
\includegraphics[width=0.32\textwidth,angle=0]{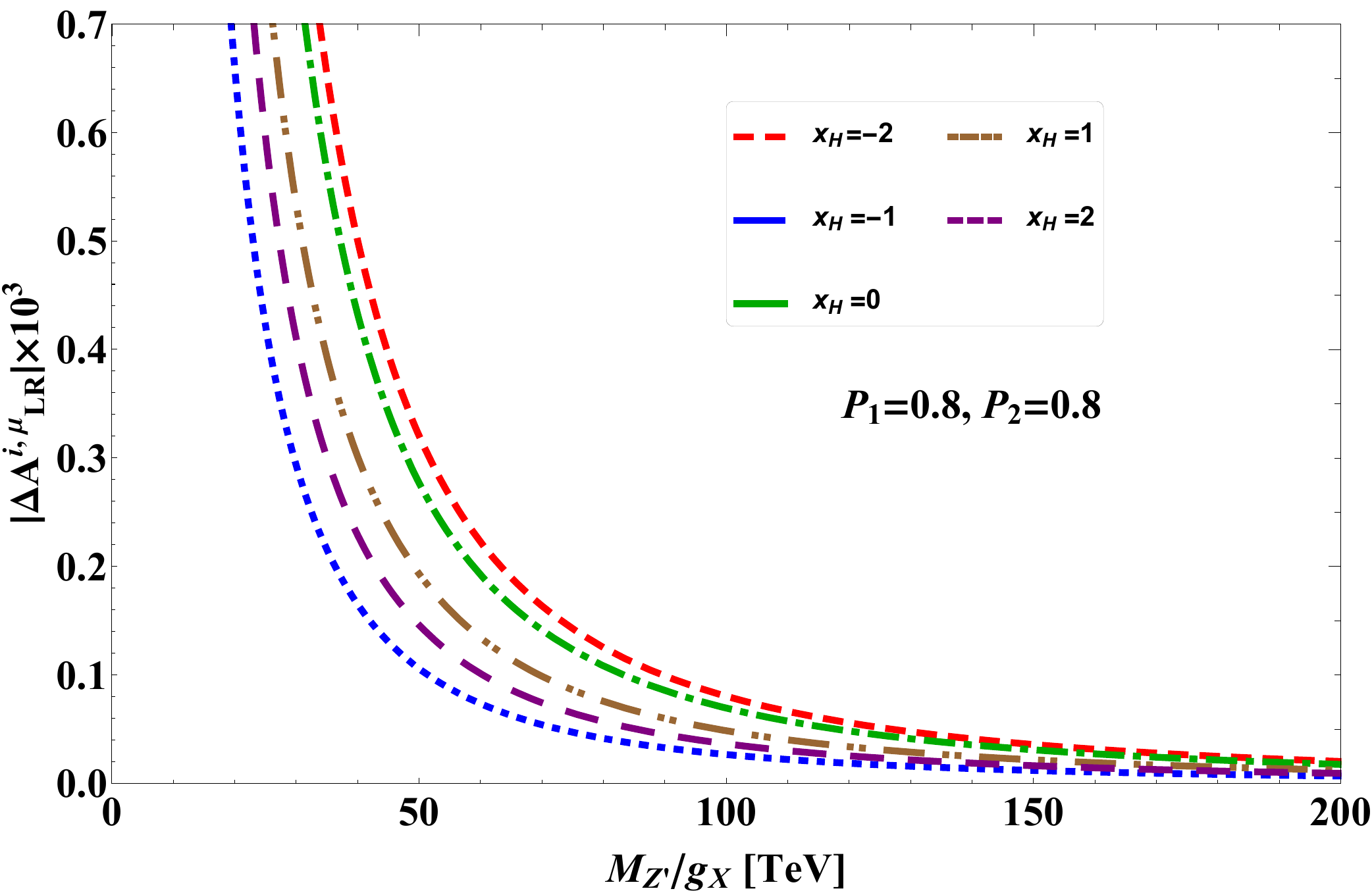}
 \caption{The total cross section for $\mu^+ \mu^+ \to \mu^+ \mu^+$ process and the corresponding deviation are shown in the upper panel from the SM for different polarization choices as a function of $\sqrt{s}$ for $M_{Z^\prime}= 7.5$ TeV and different $x_H$ for $P_{1}=P_{2}=0$, $P_{1}=P_{2}=0.8$ and $P_{1}=P_{2}=-0.8$ from left to right respectively. In the lower panel we show the differential left-right asymmetry for $\mu^+ \mu^+ \to \mu^+ \mu^+$ scattering process with respect to $\cos\theta$ in the lower left panel whereas its deviation with respect to $\cos\theta$ are given in the lower middle panel for different $x_H$. Deviation in integrated left-right asymmetry for this process as a function of $M_{Z^\prime}/g_X$ for different values of $x_H$ are shown in the lower right panel. To estimate the differential left-right asymmetry and the corresponding deviations we take $M_{Z^\prime}=7.5$ TeV. In this analysis we use $\sqrt{s}=1$ TeV in appropriate places.} 
\label{fig-sigmamu} 
\end{sidewaysfigure}
%%%%%%%%%%%%%%%%%%%%%%
The corresponding integrated scattering cross sections are shown in the upper panel of Fig.~\ref{fig-sigmamu} for different polarizations $-0.8$, $0$ and $0.8$ from left to right along with the SM contribution for $-0.99\leq \cos\theta\leq 0.99$. The integrated cross sections start deviating from the SM result from $\sqrt{s} > 1$ TeV depending on $x_H$. The integrated scattering cross section decreases with $\sqrt{s}$ showing the nature of $t-$ and $u-$channel processes, however, due to the influence of $U(1)_X$ charges and $Z^\prime$, the rates of decrease in integrated scattering cross sections are different for different polarizations depending on $x_H$. Integrated scattering cross sections for the SM process is shown by the black solid line. As a result the deviations in integrated scattering cross sections from the SM with respect to $x_H$ with the increases in $\sqrt{s}$. 

The differential left-right asymmetry of polarized cross sections is given by 
\begin{eqnarray}
A^\mu_{LR}(P_1, P_2, \cos \theta) \equiv 
 \frac{\frac{d\sigma}{d\cos\theta}(P_1, P_2) - \frac{d\sigma}{d\cos\theta}( - P_1, - P_2)}{\frac{d\sigma}{d\cos\theta}(P_1, P_2) + \frac{d\sigma}{d\cos\theta}( - P_1, - P_2)},  
\end{eqnarray}
and the deviation of the differential left-right asymmetry from the SM is given by 
\begin{eqnarray}
\Delta A^\mu_{\rm LR}\equiv \frac{[A^\mu_{\rm LR}]_{U(1)_X}}{[A^\mu_{\rm LR}]_{\rm SM}} - 1. 
\end{eqnarray}
%%%%%%%%%%%%%%%%%%%%%
The corresponding signatures are given in the lower left and middle panels of Fig.~\ref{fig-sigmamu} for different $x_H$ in the context of $U(1)_X$ scenario with $P_1=P_2=0.8$ for $M_{Z^\prime}=7.5$ TeV. The differential left right asymmetry reaches a maximum around $-0.2$ the vicinity of $\cos\theta=0$ for different values of $x_H$ whereas that for $x_H=-2$ is around $-3.5$. Due to the deviation from zero, the differential left-right asymmetry falls on both sides of $\cos\theta=0$ and showing sharp peaks for $\cos\theta=\pm1$ suggesting to consider $-1\leq\cos\theta\leq1$ while calculating integrated left-right asymmetry. We estimate the deviation in differential left-right asymmetry from the SM can reach up to $35\%$ at the vicinity of $\cos\theta=0$ for $x_H=-2$ where left handed lepton doublet do not interact with $Z^\prime$. The deviation in differential left-right asymmetry in the BSM case from the SM scenario for $x_H=-2$ slowly falls off to zero when $\cos\theta \to \pm1$. Similar behavior can be observed for $x_H=2$ but the maximum deviation may reach up to $5\%$. We notice that $\Delta\mathcal{A}_{\rm LR}$ is negative for $x_H=-1$ and $x_H=0$, however, considering $|\Delta\mathcal{A}_{\rm LR}|$ the percentage of corresponding deviations can be estimated. For example, the maximum deviations for $x_H=-1$ and $0$ can reach up to $5\%$ and $60\%$ at the vicinity of $\cos\theta=0$ which will approach zero with the limits $\cos\theta\to\pm1$. A mixed behavior will be observed for $x_H=1$ where the deviation in left-right asymmetry from the SM can be zero at the vicinity of $\cos\theta=0$ and maximum can reach up to $|\Delta\mathcal{A}_{\rm LR}|=4\%$ around $\cos\theta=\pm0.5$. In addition to that, the deviation will approach zero for $\cos\theta\to\pm1$. Apart from the dependence on $x_H$, this type of behavior occurs due to the combined effect of $t-$ and $u-$channel processes including the effects from their interference. The errors in the deviation in left-right asymmetry have been estimated using Eq.~\ref{ALR-1}. 

Finally, the integrated left-right asymmetry is given by 
\begin{eqnarray}
A^{i, \mu}_{\rm LR}
\equiv
 \frac{\int \frac{d\sigma^\mu}{d\cos\theta}(P_1, P_2)d(\cos\theta) - \int \frac{d\sigma^\mu}{d\cos\theta}( - P_1, - P_2)d(\cos\theta)}{\int \frac{d\sigma^\mu}{d\cos\theta}(P_1, P_2)d(\cos\theta) + \int \frac{d\sigma^\mu}{d\cos\theta}( - P_1, - P_2)d(\cos\theta)},  
\end{eqnarray}
and the deviation of the integrated left-right asymmetry from the SM is given by 
\begin{eqnarray}
\Delta A^{i, \mu}_{\rm LR}\equiv \frac{[A^{i, \mu}_{\rm LR}]_{U(1)_X}}{[A^{i, \mu}_{\rm LR}]_{\rm SM}} - 1
\label{LR-mu2}
\end{eqnarray}
to estimate the deviations in integrated left-right asymmetry as a function of $M_{Z^\prime}/g_X$ for different $x_H$ and these are shown in the lower right panel of Fig.~\ref{fig-sigmamu} for $P_{1}=P_{2}=0.8$\footnote{
The one-loop cross sections from the SM processes have been estimated in \cite{Bondarenko:2022kxq} for $\mu^+ \mu^+$ and $e^- e^-$ colliders. 
The relative correction between the Born process and next-to-leading order electroweak (NLO-EW) process is about $\mathcal{O}(10\%)$. 
Involving the NLO-EW cross section in $\mu^+ \mu^+$ collider with $\sqrt{s}=2$ TeV, we find ${A^{i, \mu}_{\rm LR}}_{\rm SM}^{\rm NLO-EW} = -0.0003$ whereas ${A^{i, \mu}_{\rm LR}}_{\rm SM}^{\rm Born} = -0.11$. 
The NLO-EW integrated left-right asymmetry becomes very small due to a cancellation between SM Born and one-loop contributions. 
Hence $\Delta A_{LR}^{i,\mu}$ could be enhanced.}. The $|\Delta \mathcal{A}_{\rm{LR}}^{i, \mu}|$ falls sharply for approximately $M_{Z^\prime}/g_X > 50$ TeV with respect to different $x_H$ under consideration. The bounds on $M_{Z^\prime}/g_X$ from LEP-II could be found from Tab.~\ref{tab2} for different $x_H$ which are well below $50$ TeV. Therefore $\mu$TRISTAN collider could probe a higher $U(1)_X$ breaking scale.  
%%%%%%%%%%%%%%%%%%%%%%%%%%%%%%%%%%%%
\section{Conclusions}
\label{secV}
%%%%%%%%%%%%%%%%%%%%%%%%%%%%%%%%%%%%
We study a general $U(1)$ extension of the SM to investigate the BSM effect coming from the $Z^\prime$ gauge boson in the forward dominant collisions in $\mu^+ e^-$ and $\mu^+ \mu^+$ collisions. We find that the propagators from $t-$ and $u-$channel processes involve the effect of $Z^\prime$ gauge boson manifesting different behaviors for different helicity combinations depending on general $U(1)$ charges of the leptons.
The deviations in differential and integrated scattering cross sections from the SM suggest that the influence of $Z^\prime$ could be sizable in these colliders through the $t-$ and $u-$channel processes depending on the nature of the colliders. As a result, the left-right asymmetry will be an interesting observable to test the effect of BSM physics from the $Z^\prime$ contribution. In our model due to the general $U(1)$ scenario, left and right handed fermions interact differently with $Z^\prime$ manifesting chiral scenario. Therefore we obtain sizable left-right asymmetry in the context of $\mu^+ e^-$ and $\mu^+ \mu^+$ scattering. The differential left-right asymmetry and the deviation in differential left-right asymmetry from the SM are different due to the effect of general $U(1)_X$ charges which could provide information about the interactions between $Z^\prime$ and charged leptons which could be probed in $\mu$TRISTAN experiment. The limits on the deviation in the integrated left-right asymmetry from the SM from the $\mu^+ e^-$ and $\mu^+ \mu^+$ collision can improve the bound on the general $U(1)$ breaking scale. The deviations in these analyses are more than the theoretically estimated error bars denoting that such a chiral scenario could have fascinating aspects in the forward dominant scattering. Following the sizable deviations from the SM results we could probe the nature of the interactions between $Z^\prime$ and charged leptons and also know about the breaking scale of $U(1)_X$ scenario from the $\mu$TRISTAN experiment in future. 
%%%%%%%%%%%%%%%%%%%%%%%%%%%%%%%%%%%%%%%%%%
\begin{appendix}
\section{Error analyses}
\begin{widetext}
The error bar of Left-Right asymmetry at $\theta=\theta_0$ is given by 
\begin{eqnarray}
\delta \Delta A^i_{LR}(\cos\theta_0) = 
2\frac{\sqrt{N^i_{LR}(\cos\theta_0) N^i_{RL}(\cos\theta_0)}\left(\sqrt{N^i_{LR}(\cos\theta_0)} + \sqrt{N^i_{RL}(\cos\theta_0)}\right)}{(N^i_{LR}(\cos\theta_0)+N^i_{RL}(\cos\theta_0))^2}, 
\label{ALR-1}
\end{eqnarray}
where $i$ is e or $\mu$, 
\begin{eqnarray}
N^i_{LR}(\cos\theta_0)\equiv {\cal L}_{int}\int_{\cos\theta_0-0.05}^{\cos\theta_0+0.05} \frac{d\sigma^i}{d\cos\theta}(P_1, P_2, \cos \theta)
d \cos\theta, 
\\
N^i_{RL}(\cos\theta_0)\equiv {\cal L}_{int}\int_{\cos\theta_0-0.05}^{\cos\theta_0+0.05} \frac{d\sigma^i}{d\cos\theta}(-P_1, -P_2, \cos \theta)
d \cos\theta, 
\end{eqnarray}
and ${\cal L}_{int}$ is the integrated luminosity. 
\end{widetext}
\end{appendix}
%\vspace{-0.398in}
\bibliographystyle{utphys}
\bibliography{bibliography}
\end{document}